\documentclass[useAMS,usenatbib]{mn2e}
\usepackage{graphicx}

\usepackage[applemac]{inputenc}
\usepackage{amsmath}

\title[Non-Gaussian Signatures in the five-year WMAP data]
      {Non-Gaussian Signatures in the five-year WMAP data as identified with isotropic scaling indices}

\author[G. Rossmanith, C. Räth, A. J. Banday and G. Morfill]
       {G. ~Rossmanith$^1$\thanks{E-mail: rossmanith@mpe.mpg.de}, C. ~Räth$^1$, A. J. ~Banday$^{2,3}$ and G. ~Morfill$^1$  \\
        $^1$ Max-Planck-Institut für extraterrestrische Physik, Giessenbachstr. 1, D-85748 Garching, Germany \\
        $^2$ Centre d’Etude Spatiale des Rayonnements, 9, Av du Colonel Roche, 31028 Toulouse, France \\
        $^3$ Max-Planck-Institut für Astrophysik, Karl-Schwarzschild-Str. 1, D-85741 Garching, Germany}

\begin{document}

\date{Accepted ...
      Received .....;
      in original form ....}
\pagerange{\pageref{firstpage}--\pageref{lastpage}} \pubyear{2009}

\maketitle

\label{firstpage}

\begin{abstract}
We continue the analysis of non-Gaussianities in the CMB by means of the scaling index method (SIM, Räth, Schuecker \& Banday 2007) by applying this method on the single Q-, V-, W-bands and the co-added VW-band of the 5-year data of the \textit{Wilkinson Microwave Anisotropy Probe (WMAP)}. We compare each of the results with 1000 Monte Carlo simulations mimicing the Gaussian properties of the best fit $\Lambda CDM$-model. Based on the scaling indices, scale-dependent empirical probability distributions, moments of these distributions and $\chi^2$-combinations of them are calculated, obtaining similar results as in the former analysis of the 3-year data: We derive evidence for non-Gaussianity with a probability of up to 97.3\% for the mean when regarding the KQ75-masked full sky and summing up over all considered length scales by means of a diagonal $\chi^2$-statistics. Looking at only the northern or southern hemisphere of the galactic coordinate system, we obtain up to 98.5\% or 96.6\%, respectively. For the standard deviation, these results appear as 95.6\% for the full sky (99.7\% north, 89.4\% south) and for a $\chi^2$-combination of both measurements as 97.4\% (99.1\% north, 95.5\% south). We obtain larger deviations from Gaussianity when looking at seperate scale lengths. By performing an analysis of rotated hemispheres, we detect an obvious asymmetry in the data. In addition to these investigations, we present a method of filling the mask with Gaussian noise to eliminate boundary effects caused by the mask. With the help of this technique, we identify several local features on the map, of which the most significant one turns out to be the well-known \textit{cold spot}. When excluding all these spots from the analysis, the deviation from Gaussianity increases, which shows that the discovered local anomalies are not the reason of the global detection of non-Gaussianity, but actually were damping the deviations on average. Our analyses per band and per year suggest, however, that it is very unlikely that the detected anomalies are due to foreground effects.
\end{abstract}



\section{Introduction}

The \textit{Wilkinson Microwave Anisotropy Probe (WMAP)} satellite, launched in June 2001, measures the temperature anisotropy of the cosmic microwave background (CMB) radiation with surpassing accuracy, hence providing the best insight on the beginnings of our universe until now. From the first data release on, many investigations were made concerning the Gaussianity of the CMB, since such analyses give information about the nature of the primordial density fluctuations, which are the seeds of those temperature anisotropies. The statistical properties of the density fluctuations are again an important observable for testing cosmological models, especially models of inflation. Standard inflationary models predict the temperature fluctuations of the CMB to be a Gaussian random field which is isotropic and homogenous \citep{guth81a, linde82a, albrecht82a}. Still, there also exist more complex models that allow non-Gaussianity in a scale-independend \citep{linde97a, peebles97a, bernardeau02a, bartolo02a, lyth02a, lyth03a, acquaviva03a} or in a scale-dependend way \citep{garriga99a, armendariz99a}. For a detailed overview of the different models and a more specific survey on scale-dependend ones, see \citet{bartolo04a} and \citet{loverde08a}, respectively, as well as enclosed references. In addition, topological defects like cosmic strings can induce local non-Gaussianities and influence the power spectrum \citep{kaiser84a, bouchet88a, turok90a, turok96a, jeannerot03a}. Considering this plethora of possible physical mechanisms, which may induce non-Gaussianity, studies of Gaussianity of the CMB are strongly required for testing predictions of fundamental physical theories. By comparing the results with theoretical predictions, we can evaluate which model e.g. of inflation can be accepted or rejected. \par
Non-Gaussianity implies the presence of any higher order correlations. Therefore, a concrete description of the characteristics of non-Gaussianity is not possible, and one has to state that it can occur in various forms. One can carry out a global analysis to search for deviations from Gaussianity \citep{komatsu03a, komatsu09a, chiang03a, chiang07a, coles04a, odwyer04a, eriksen07a, smith09a, rudjord09a}. But one can also concentrate on more specific investigations (this is most often performed in addition to a general analysis), whereas we want to point out the following two: \par
Investigations concerning \textit{asymmetries} in the CMB data were accomplished with linear \citep{eriksen04a, hansen04b, hansen08a, bonaldi07a, bernui08a, bernui08b, hoftuft09a} as well as non-linear methods \citep{eriksen04a, eriksen04b, park04a, vielva04a, hansen04a, land05a, pietrobon08a, raeth09a}. With those methods, studies of the differences between the northern and southern hemisphere of the galactic coordinate system, naturally given by the absent region of the outmasked galactic plane, as well as a search for a preferred direction of maximum asymmetry were performed. In almost all investigations, significant asymmetries between the north and the south were detected. Thereby, it depended on the type of analysis, which hemisphere featured the larger deviations from Gaussianity and which hemisphere agreed better with the standard model. The preferred direction of maximal disparity was in most investigations found to lie close to the ecliptic axis. \par
\textit{Local features} are another particular form of non-Gaussianity being of growing importance, e.g. for the search of topological defects like cosmic strings. Since the first detection of the famous \textit{cold spot} by \citet{vielva04a}, many investigations tried to find new, or re-detect known spots with various methods \citep{mukherjee04a, cayon05a, cruz05a, cruz07b, mcewen05a, mcewen06a, mcewen08a, vielva07a, pietrobon08a, gurzadyan08a, gurzadyan09a}. In doing so, several significant spots have been detected up to now. \par
In \citet{raeth07a}, all mentioned investigations were accomplished by applying, for the first time, the scaling index method on the WMAP 3-year data. In this paper, we continue these analyses by applying the scaling index method on the WMAP 5-year observations. We search for global non-Gaussianities and asymmetries in the data and use a modified approach to detect local features. \par
This paper is structured as follows: In Section 2 we present the preprocessing of the WMAP data and the modality of creating the simulations. In Section 3, the scaling index method is introduced as well as a technique to cope with boundary effects. With these requisites, we are ready to perform our calculations, whose results are presented in Section 4. In this chapter, we first discuss the global investigations as well as asymmetries and focus on local features later on. All these findings are summarised in Section 5. Finally, we draw our conclusions in Section 6.


\section{WMAP Data and Simulations}
\label{Kapitel2}

For our investigations we use the Q-, V- and W-band five-year-data of the WMAP-satellite as it is provided by the WMAP-Team\footnote{http://lambda.gsfc.nasa.gov\label{refinote}}. We work with the foreground-reduced maps, which use the Foreground Template Model proposed in \citet{hinshaw07a} and \citet{page07a} for foreground reduction. To obtain a co-added VW-map as well as single V-, W- and Q-maps, we accumulate the differencing assemblies Q1, Q2, V1, V2, W1, W2, W3, W4 via a noise-weighted sum \citep{bennett03a}:
\begin{equation} \label{GewSum}
T(\theta, \phi) = \frac{\sum_{i \in \mathcal{A}} T_i(\theta,\phi)/ \sigma^2_{0,i} } {\sum_{i \in \mathcal{A}} 1/ \sigma^2_{0,i}}
\end{equation}
In this equation, $\mathcal{A}$ characterises the set of required assemblies, e.g. for the co-added VW-map $\mathcal{A} = \lbrace V1,V2,W1,W2,W3,W4 \rbrace$. The parameters $\theta$ and $\phi$ correspond to the co-latitude and the longitude on the sphere, while the five-year noise per observation of the different assemblies, given by \citet{hinshaw09a}, is denoted by $\sigma_0$. \par
\begin{figure*}
\centering
\includegraphics[width=8cm, keepaspectratio=true, ]{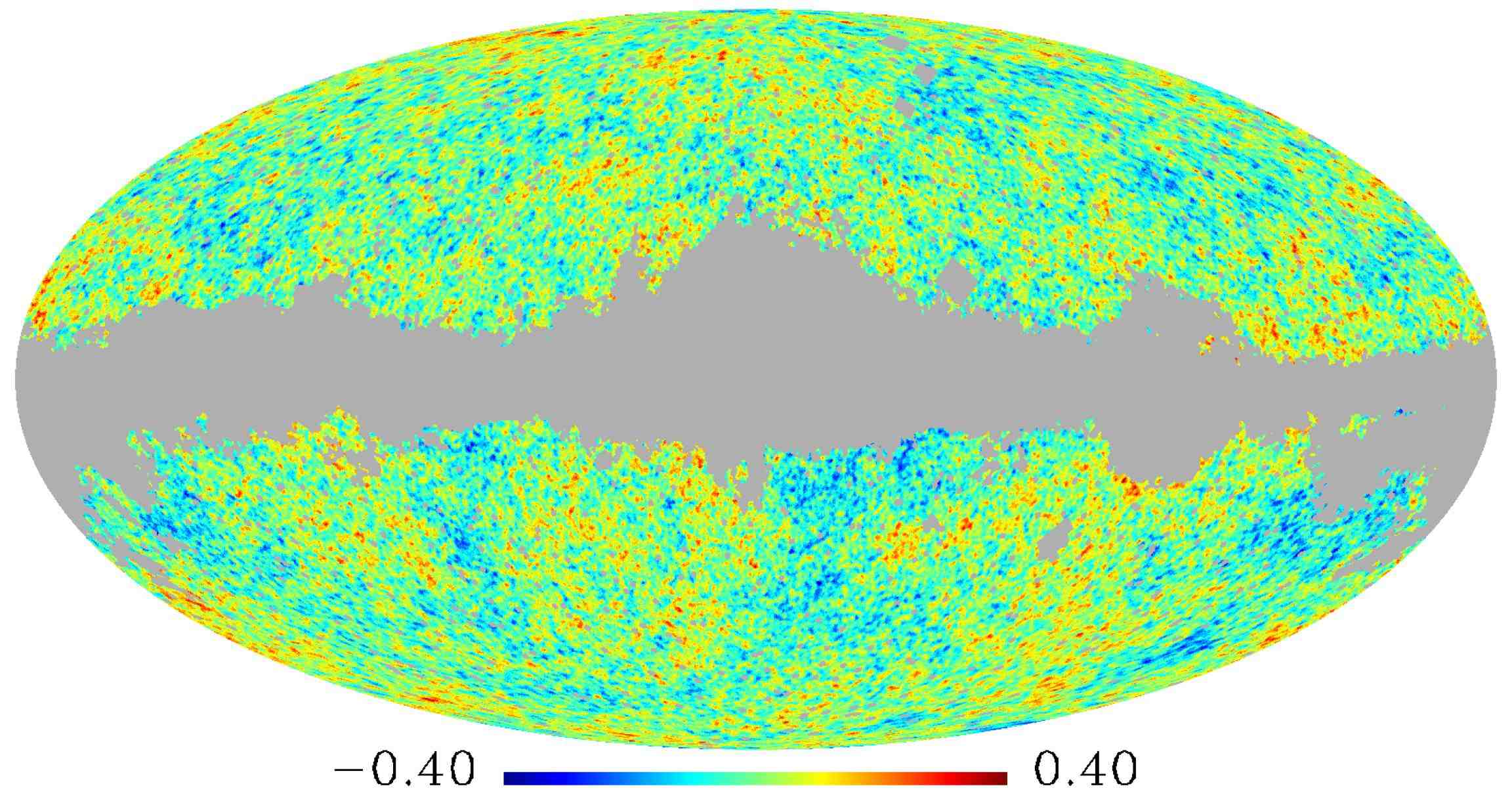}
\includegraphics[width=8cm, keepaspectratio=true, ]{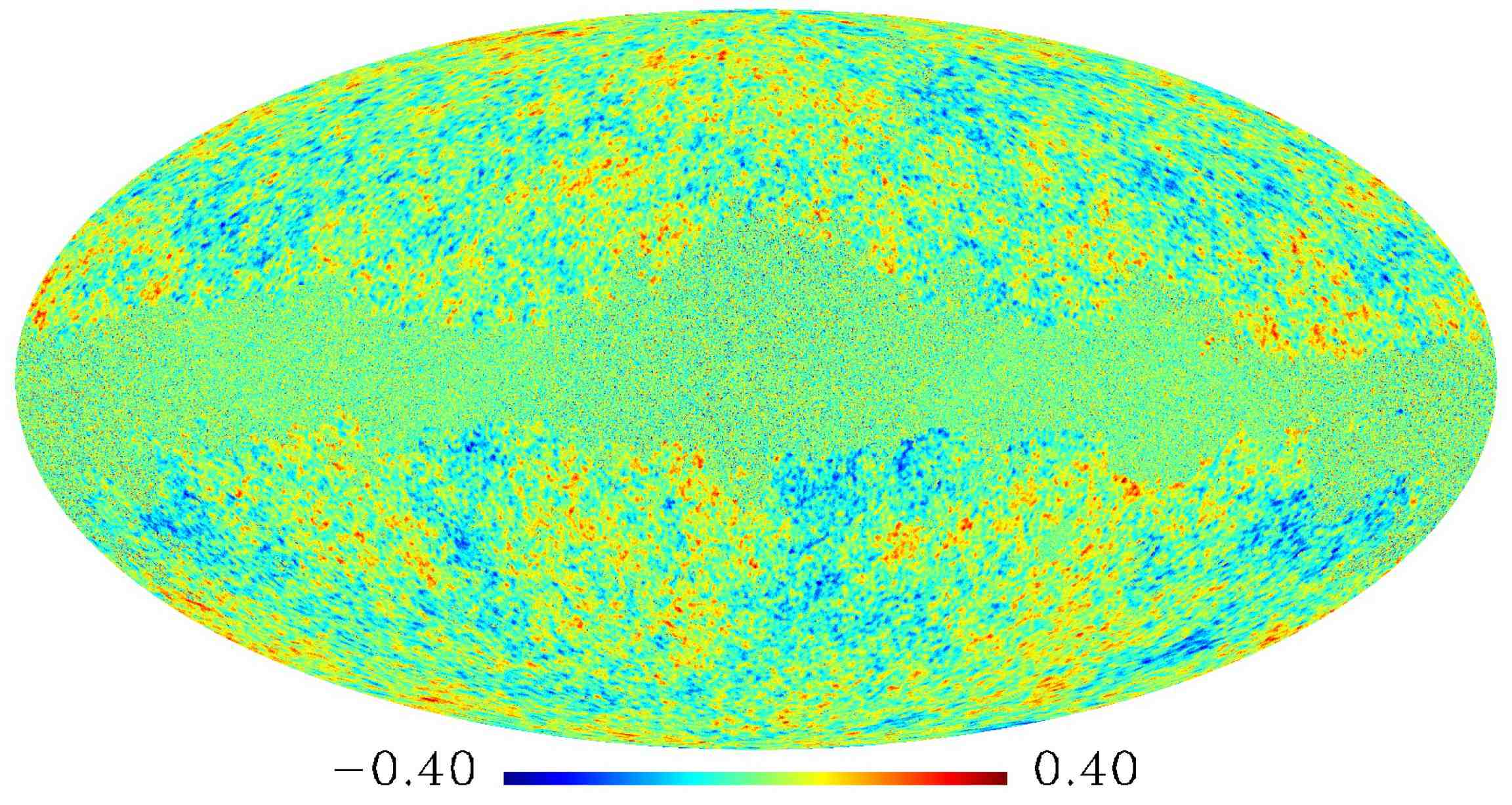}
\includegraphics[width=8cm, keepaspectratio=true, ]{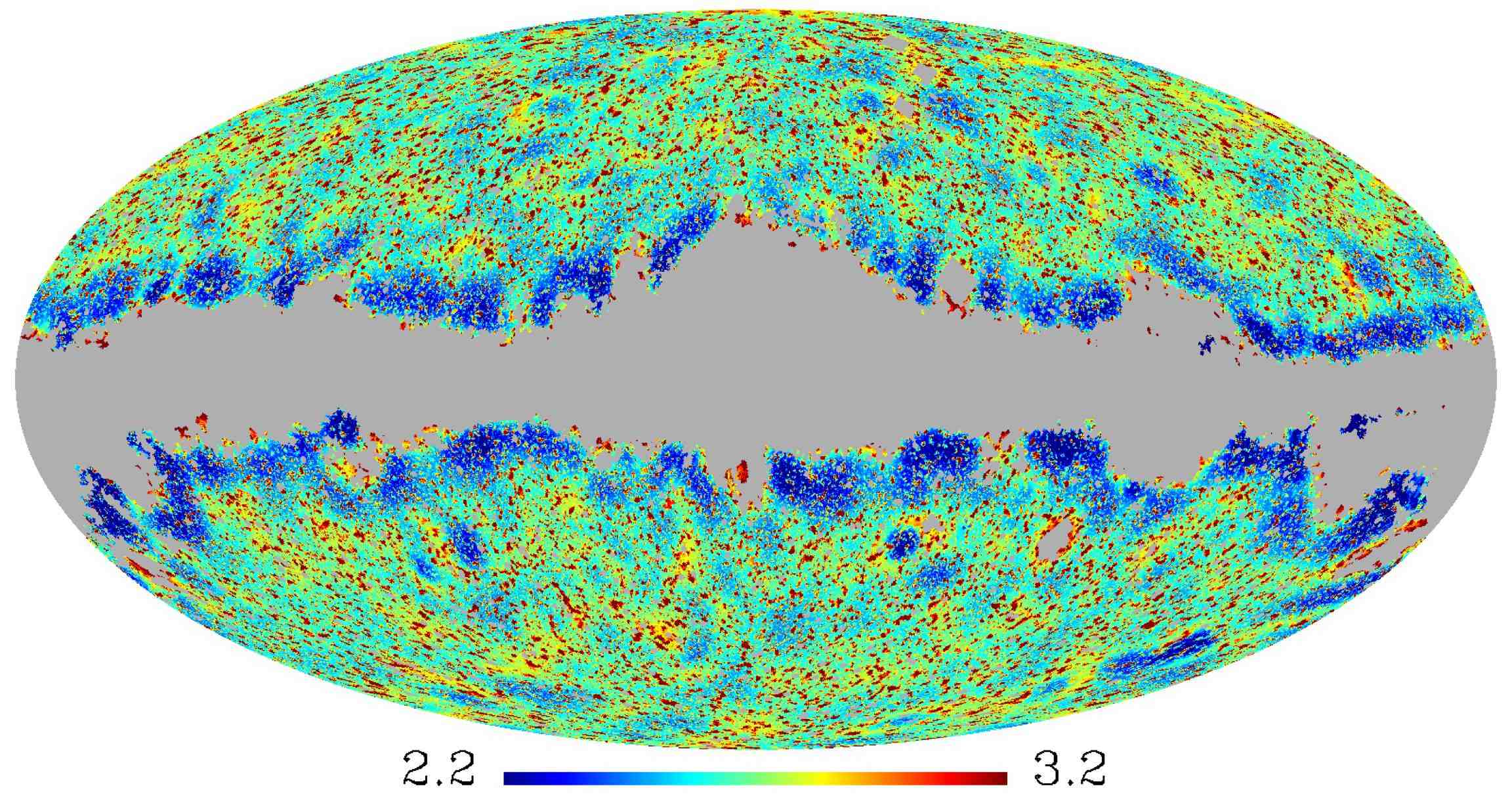}
\includegraphics[width=8cm, keepaspectratio=true, ]{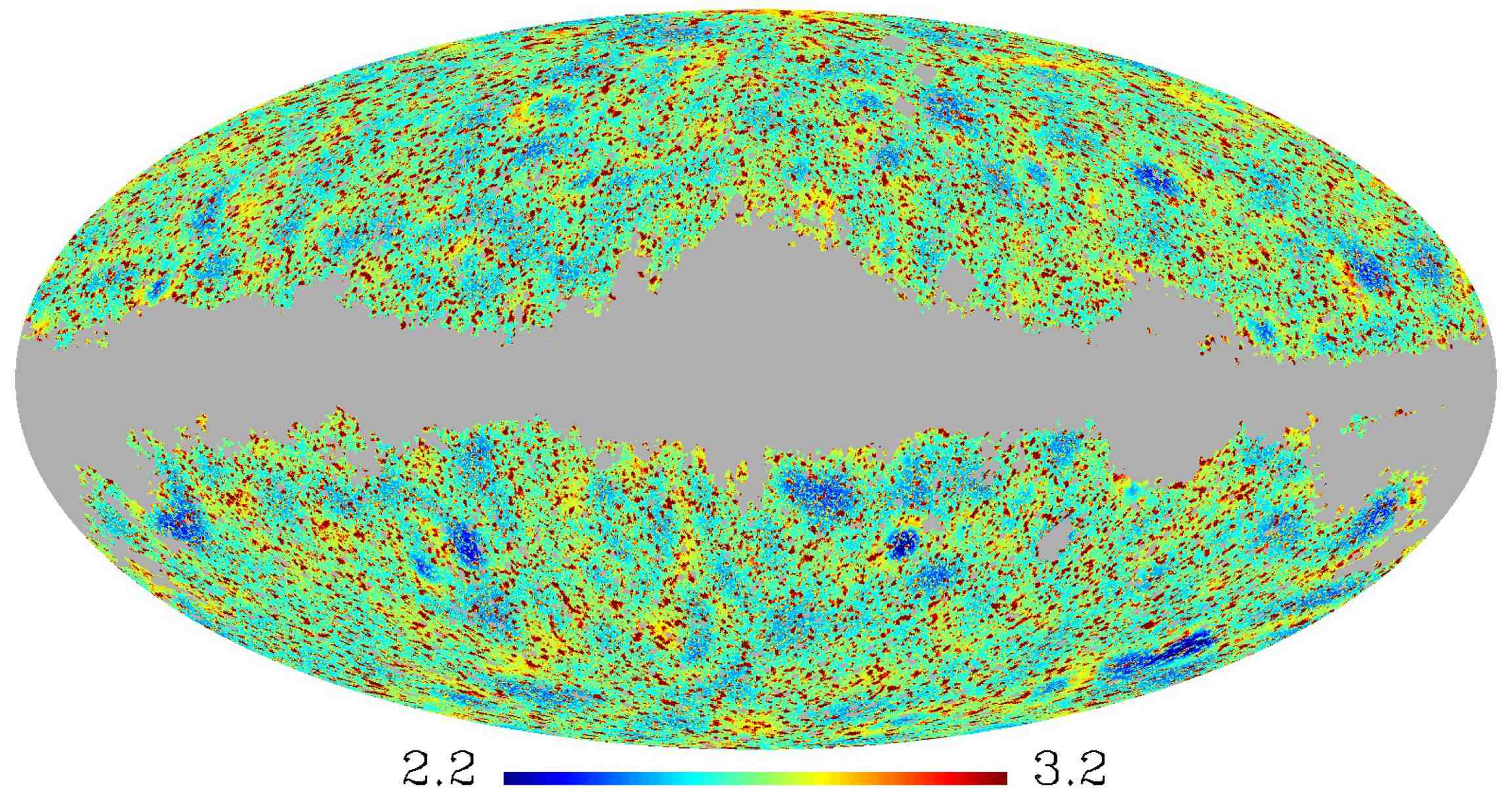}
\caption{The two plots on the left hand side illustrate the original 5-year WMAP-map of the co-added VW-band (above) and the related colour-coded $\alpha$-response (below). The equivalent plots for the mask-filling technique are arranged on the right hand side. These maps (and all following ones) are shown in a conventional scheme, namely the Mollweide projection in the Galactic reference frame with the Galactic Centre at the centre of the image and the longitude increasing from there to the left-hand side.} \label{fig1:dreimaldrei}
\end{figure*}

We decrease the resolution of the maps to 786432 pixels, which equals to  $N_{side} =  256$ in the employed HEALPix-software\footnote{http://healpix.jpl.nasa.gov} \citep{gorski05a} and cut out the heavily foreground-affected parts of the sky using the KQ75-mask \citep{gold09a}, which has to be downgraded as well. We choose a conservative downgrading of the mask by taking only all pixels at $N_{side} =  256$ that do completely consist of non-mask-pixels at $N_{side} =  512$. All downgraded pixels at $N_{side} =  256$, for which one or more pixels at $N_{side} =  512$ belonged to the KQ75-mask, are considered to be part of the downgraded mask as well. In doing so, $28.4\%$ of the sky is removed (see upper left part of figure \ref{fig1:dreimaldrei}). Finally, we remove the residual monopol and dipol by means of the appropriate HEALPix routine applied to the unmasked pixels only. \par

To accomplish a test of non-Gaussianity, we also need simulations of Gaussian random fields. We create 1000 simulations for every band and proceed hereby as follows: We take the best fit $\Lambda CDM$ power spectrum $C_l$, derived from the WMAP 5-year data only, and the respective window function for each differencing assembly (Q1-Q2, V1-V2, W1-W4), as again made available on the LAMBDA-website$^1$. With these requisites, we can create Gaussian random fields mimicing the Gaussian properties of the best fit $\Lambda CDM$-model and including the WMAP-specific beam properties by convolving the $C_l$´s with the window function. For every assembly, we add Gaussian noise to these maps with a particular variance for every pixel of the sphere. This variance depends on the number of observations $N_i(\theta,\phi)$ in the respective direction and the noise dispersion per observation, $\sigma_{0,i}$. After this procedure, we summarize the Q-, V- and W-bands and the co-added VW-band using equation (\ref{GewSum}), decrease the resolution to $N_{side} =  256$, cut out the KQ75-mask and remove the residual monopol and dipol, just as we did with the WMAP-data.

  
\section{Weighted Scaling Index Method}

\subsection{Formalism}
\label{Formalism}

We perform our investigations using the scaling index method (SIM) \citep{raeth02a, raeth03a}, which enables a characterisation of the structure of a given data set. It has already been used in time series analysis of active galactic nuclei (AGN) \citep{gliozzi02a, gliozzi06a} as well as in structure analysis for 2D and 3D image data, e.g. in \citet{jamitzky01a, monetti03a, raeth08a}. In the following, we only present a short overview of the calculation of scaling indices. For a more detailed formalism of using the SIM in CMB analysis we refer to \citet{raeth07a}. \par

\begin{table}
\begin{tabular}{lccccc}
\hline \hline
Radius & 0.025 & 0.050 & 0.075 & 0.100 & 0.125 \\
$[\ell_1,\ell_2]$ & [83,387] & [41,193] & [28,129] & [21,97] & [17,77] \\ \hline
Radius & 0.150 & 0.175 & 0.200 & 0.225 & 0.250 \\
$[\ell_1,\ell_2]$ & [14,65] & [12,55] & [10,48] & [9,43] & [8,39] \\ \hline \hline
\end{tabular}
\caption{The angular scales corresponding to the position of the $90\%$ ($\ell_1$) and the $10\%$ ($\ell_2$) weighting in the scaling index formula when using a given scale parameter $r$.} \label{TableNull}
\end{table}

The fluctuations of the temperature maps are characterized by the values of the pixelised sky of a sphere with radius $R$. We transform this representation to variations in the radial direction around the sphere by applying a jitter depending on the intensity of the fluctuation. Thereby we obtain a point-distribution in the three-dimensional space. Thus, given $N_{pix}$ as the number of pixels on the sphere, the value of every pixel $(\theta_i,\phi_i)$, $i = 1,...,N_{pix}$ corresponds to a vector $\vec{p}_i$ in the three-dimensional space. We then define for every point $\vec{p}_i$ its scaling index by
\begin{equation} \label{AlphaSpeziell}
\alpha(\vec{p}_i, r) = \frac{ \sum_{j=1}^{N_{pix}}  2\left(\frac{d_{ij}}{r}\right)^2 e^{-\left(\frac{d_{ij}}{r}\right)^2} }{ \sum_{j=1}^{N_{pix}} e^{-\left(\frac{d_{ij}}{r}\right)^2} }
\end{equation}
where $d_{ij}$ denotes the euclidian distance measure
\begin{equation*} \label{Euclid}
d_{ij} = \| \vec{p}_i - \vec{p}_j \|_2
\end{equation*}
between the points $\vec{p}_i$ and $\vec{p}_j$, while $r$ characterizes a scale parameter. This parameter does not draw a clear-cut line between the pixels that are included in the calculations and those that are excluded; it rather influences how each single pixel is taken into consideration for the calculation, in relation to its distance from the center pixel: For lower $r$, only the closest pixels are important in the calculation of $\alpha(\vec{p}_i, r)$, whereas for larger $r$, the farther distant pixels are considered as well, even though with a still lower weight than the close pixels. In our study, we use the ten scales $r_i = 0.025, 0.05, ..., 0.25$, $i = 1, 2, ..., 10$ and the radius $R=2$ for the sphere. Table \ref{TableNull} shows for each radius $r$ the corresponding angular scales at the position of the 90\% and the 10\% weighting, thus giving an estimate on how the r-values relate to $\ell$-bands in Fourier space. \par
The value of $\alpha$ characterises the structural components of a point distribution. For example, points in a cluster-like, filamentary or sheet-like structure lead to $\alpha \approx 0$, $\alpha \approx 1$ or $\alpha \approx 2$, respectively. A uniform distribution of points results in $\alpha \approx 3$, while points in underdense regions in the vicinity of point-like structures, ﬁlaments or walls have $\alpha > 3$. \par
On the basis of these scale-dependent $\alpha$-values, we compute simple measures such as moments and empirical probability distributions. We make use of the following scale-dependend statistics, namely the mean, the standard deviation and a diagonal $\chi^2$-statistics, to compare the results of the original WMAP data with the results of the simulations:
\begin{align} \label{MeanStdevChi}
\langle \alpha(r_k) \rangle \ &= \ \frac{1}{N} \sum_{j=1}^{N} \alpha(\vec{p}_j, r_k) \\
\sigma_{\alpha(r_k)} \ &= \ \left( \frac{1}{N-1} \sum_{j=1}^{N} \left[ \alpha(\vec{p}_j, r_k) - \langle \alpha(r_k) \rangle \right] ^2 \right)^{1/2} \\
\chi^2_{\alpha(r_k)} \ &= \ \sum_{i=1}^{2} \left[ \frac{M_i(r_k) - \langle M_i(r_k) \rangle }{\sigma_{M_i(r_k)}} \right] ^2 \label{Zeile3}
\end{align}
where $M_1(r_k) = \langle \alpha(r_k) \rangle$, $M_2(r_k) = \sigma_{\alpha(r_k)}$ and $N$ denotes the number of pixels in consideration. For all analyses we will only consider the non-masked pixels of the full sky or of (rotated) hemispheres, as it will be outlined in Section \ref{VierEins}. Note that we follow the reasoning of \citet{eriksen04b} and choose a diagonal and not the full $\chi^2$-statistics involving the inverted cross-correlation matrix, because also in our case the moments are highly correlated leading to high values in the off-diagonal elements of the cross-correlation matrix. Therefore, the matrix would converge very slowly and numerical stability would not be given. If however the chosen model is a proper description of the data, \textit{any} combination of measures should yield statistically the same values for the observations and the simulations. \par

\begin{figure}
\centering
\includegraphics[width=8cm, keepaspectratio=true, ]{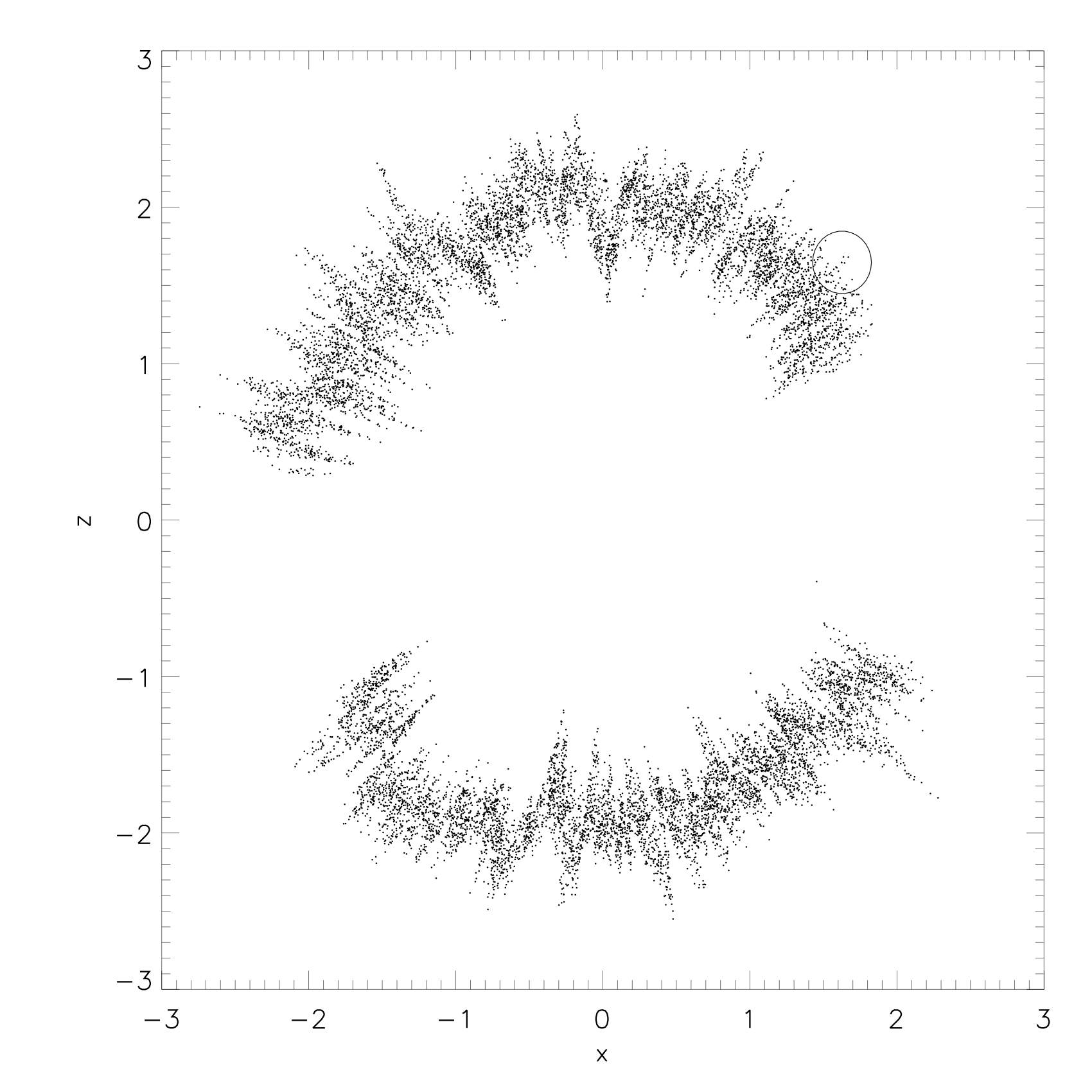}
\includegraphics[width=8cm, keepaspectratio=true, ]{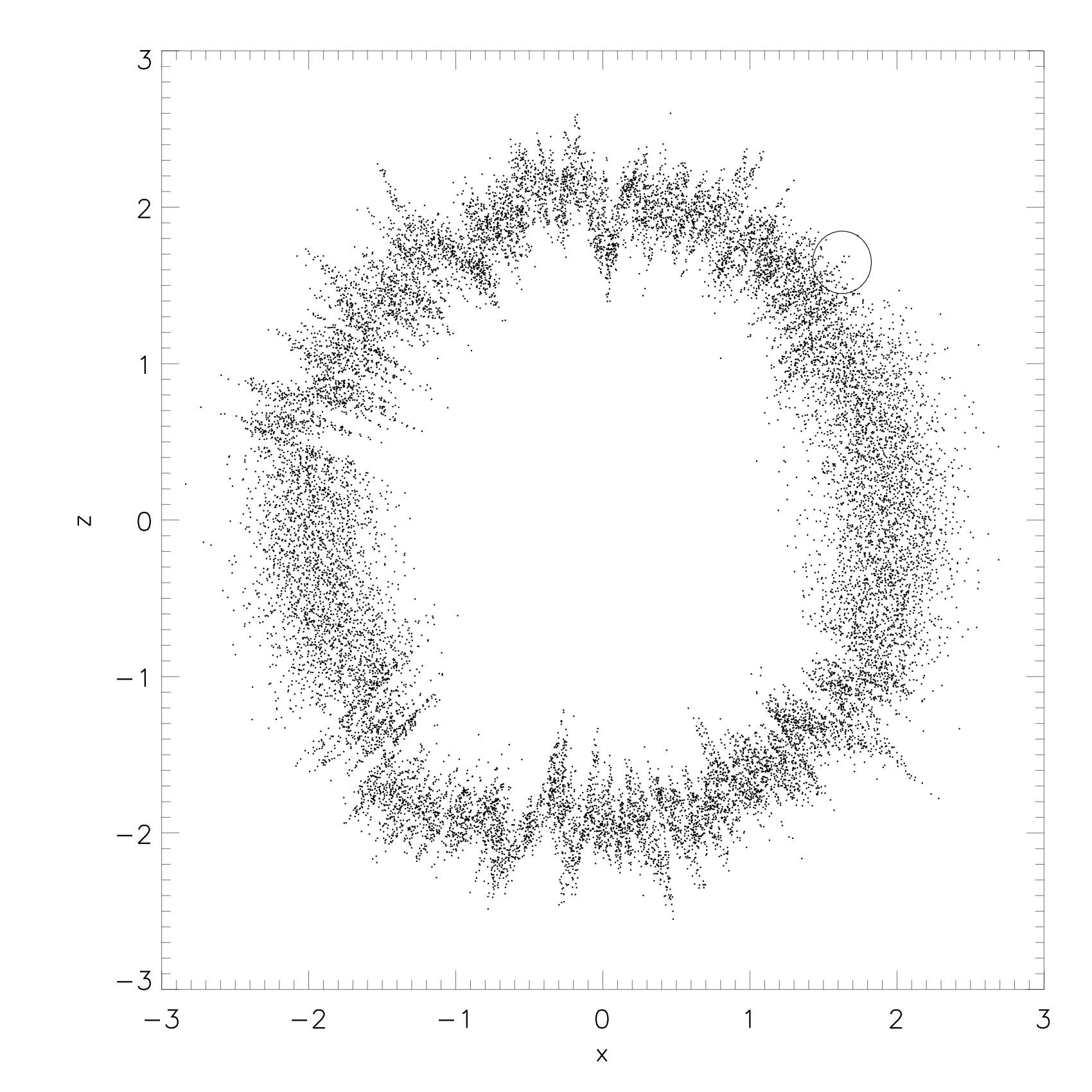}
\caption{A slice of the three-dimensional representation of the VW-band WMAP data, illustrated as a $x$,$z$-projection of all points with $\left|y\right|<0.05$. The upper plot illustrates the original, the lower the mask-filling method. The black circles indicate the scaling range $r=0.2$.} \label{fig2:jitter}
\end{figure}

To obtain scale-independent variables as well, we also use three diagonal $\chi^2$-statistics, derived from $\langle \alpha \rangle$ and $\sigma_{\alpha}$, which sum over all utilised lenght scales $r$:
\begin{align} \label{ChiSummed}
\chi^2_{\langle \alpha \rangle} \ &= \ \sum_{k=1}^{N_r} \left[ \frac{M_1(r_k) - \langle M_1(r_k) \rangle }{\sigma_{M_1(r_k)}} \right] ^2 \\
\chi^2_{\sigma_{\alpha}} \ &= \ \sum_{k=1}^{N_r} \left[ \frac{M_2(r_k) - \langle M_2(r_k) \rangle }{\sigma_{M_2(r_k)}} \right] ^2 \\
\chi^2_{\langle \alpha \rangle,\sigma_{\alpha}} &= \ \sum_{i=1}^{2} \sum_{k=1}^{N_r} \left[ \frac{M_i(r_k) - \langle M_i(r_k) \rangle }{\sigma_{M_i(r_k)}} \right] ^2 \label{ZeileC}
\end{align}
Hereby denotes $M_1(r_k) = \ \langle \alpha(r_k) \rangle $ and $M_2(r_k) = \sigma_{\alpha(r_k)}$. The number of different scale parameters $r$ is named $N_r$. Throughout all subsequent investigations, $N_r$ equals ten. \par
Finally, to be able to access the degree of difference between the data and the simulations and hence a degree of the non-Gaussianity of the data, we use the $\sigma$-normalised deviation of the WMAP results of the above-mentioned statistics:
\begin{equation} \label{Signifikanz}
S = \frac{M - \langle M \rangle}{\sigma_M}
\end{equation}
where in this case $M$ refers to one of the variables defined in equation (\ref{MeanStdevChi}) to (\ref{ZeileC}) respectively. $M$ itself is calculated by using the WMAP data, while its moments result from the simulations. Note that we pass on the absolute value in this general definition to obtain positive as well as negative deviation. Although we will use the absolute value in the global investigations, the sectioning into positive and negative deviation is useful for the analysis of north-south asymmetry by means of rotated hemispheres in chapter \ref{VierEins}. It also allows a better interpretation of the character of difference, since e.g. a higher mean of the scaling indices implies a more 'unstructured' arrangement of the 'pixel cloud' and vice versa. Similarily, a higher standard deviation of the indices indicates a larger structural variability. \par
In the tables, we also included the fraction $p$ of the simulations that have higher (lower) values than the data in terms of the respective calculated statistics. This percentage corresponds to a empirical significance level of the null hypothesis that the observation belongs to a Gaussian Monte Carlo ensemble.


\subsection{Coping with boundary effects}
\label{Copingboundary}

The regions in the direction of the galactic plane as well as many small spots all over the WMAP map are masked out since they represent heavily foreground-affected areas which would not allow a reasonable analysis of the intrinsic background fluctuations. But this operation spoils the results of the scaling index method: Instead of a more or less uniform distribution, the $\alpha$-values in the regions around the mask now detect a sharp boundary with no points in the masked area, into which the scaling regions extends (see figure \ref{fig2:jitter}). This results in lower values of $\alpha$. The effect can clearly be seen in the $\alpha$-response of the masked VW-band WMAP-data in the lower left corner of figure \ref{fig1:dreimaldrei}. A solution to this problem is to \textit{fill} the masked areas with suitable values, that prevent the low outcome at the edges of the mask. We accomplish this by filling in (nearly) white Gaussian noise with adjusted parameters. This is performed by applying the following two steps: \par
At first, we fill the masked regions with Gaussian noise, whose standard deviation for each pixel corresponds to the pixel noise made available on the LAMBDA-website\footnote{http://lambda.gsfc.nasa.gov}:
\[ T_{mask}^{\ast}(\theta, \phi) \sim \mathcal{N}( 0 , \sigma_{(\theta, \phi)}^2 ) \]
Here, $\sigma_{(\theta, \phi)}$ denotes the pixel noise of the pixel which is located in the direction $(\theta, \phi)$. Then, we scale the expectation value and the variance as a whole to the empirical mean $\mu_{rem}$ and variance $\sigma_{rem}^2$ of the remaining regions of the original temperature map:
\[ T_{mask}(\theta, \phi) = \frac{\sigma_{rem}^2}{\sigma_{mask}^2} \ T_{mask}^{\ast}(\theta, \phi) + \mu_{rem} \]
with
\begin{align*} \label{Fuellformel1_zusatz}
\mu_{rem} \ \ &= \ \frac{1}{N_\mathcal{R}} \sum_{(\theta, \phi) \in \mathcal{R}} T(\theta, \phi)  \\
\sigma_{rem}^2 \ \ &= \ \frac{1}{N_\mathcal{R}-1}\sum_{(\theta, \phi) \in \mathcal{R}} (T(\theta, \phi)-  \mu_{rem})^2  \\
\sigma_{mask}^2 \ &= \ \frac{1}{N_\mathcal{M}-1}\sum_{(\theta, \phi) \in \mathcal{M}} T_{mask}^{\ast}(\theta, \phi)^2
\end{align*}
where $\mathcal{R}$ and $\mathcal{M}$ stand for the non-masked and masked region of the map respectively, and $N_\mathcal{R}$ as well as $N_\mathcal{M}$ denote their number of pixels. Thus, we filled the mask with (nearly) white Gaussian noise whose mean and standard deviation equal the respective terms of the remaining map, whereby the spatial noise patterns are preserved. \par
With this filling technique, we obtain a complemented data set instead of just excluding the masked regions. Figure \ref{fig2:jitter} shows a slice of the three-dimensional representation of the temperature fluctuations of both techniques, representing a 2D projection of the 3D point distribution used for the calculation of the scaling indices. The center region of the filled sphere now highly resembles the appearence of the remaining area, although a more uniform arrangement is visible. In the form of a mollweide projection, the filling method as well as the corresponding $\alpha$-response are displayed in the right column of figure \ref{fig1:dreimaldrei}. \par

\begin{figure}
\centering
\includegraphics[width=8cm, keepaspectratio=true]{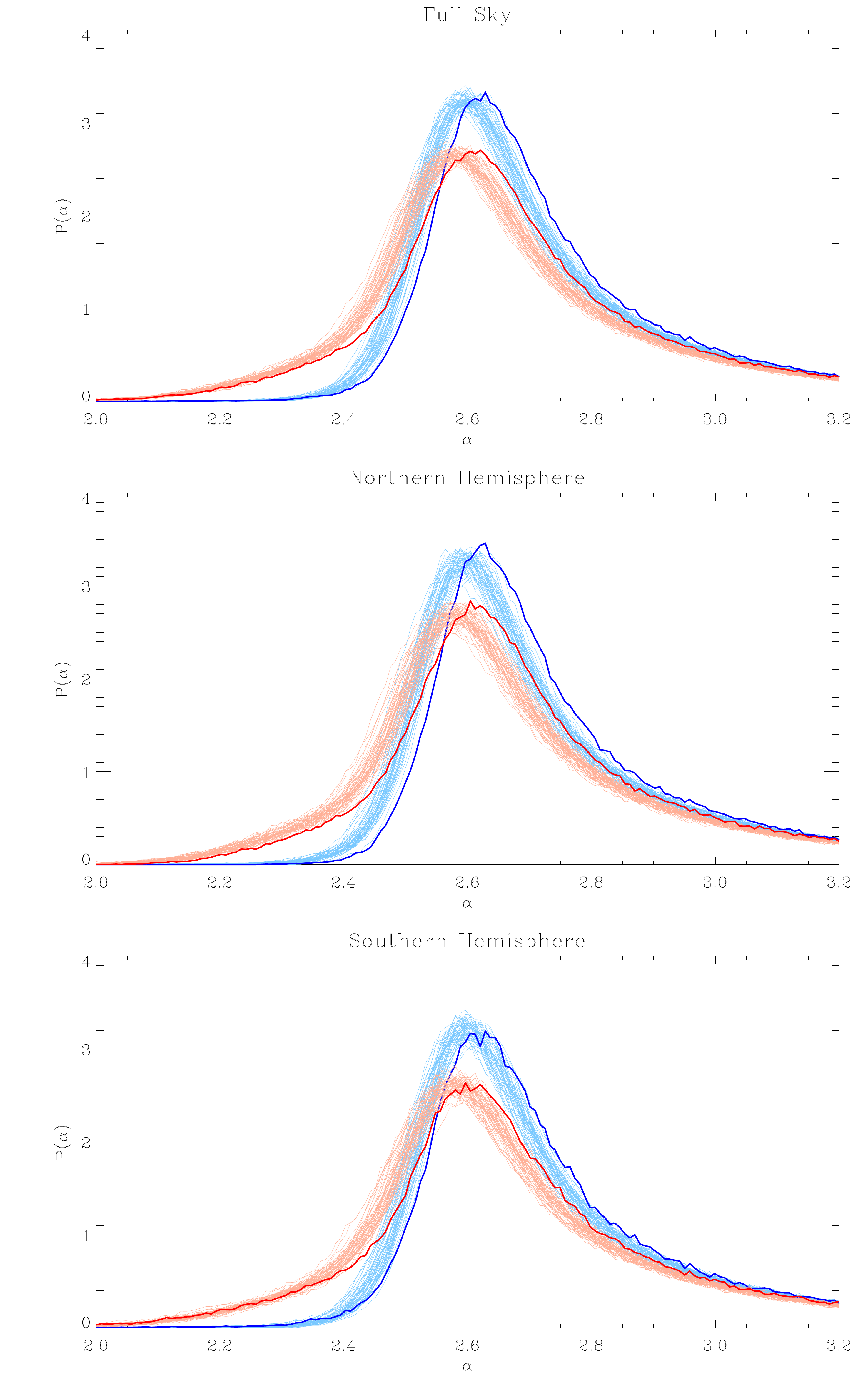}
\caption{The probability distributions $P(\alpha)$ of the scaling indices for the WMAP data (dark lines) and for 50 simulations (fainter lines) by using the scale parameter $r=0.2$, computed for the original (red) and the mask filling method (blue). The upper histogram shows the distribution of the full sky data set, while the middle and the lower ones show the distribution of the northern and southern hemisphere respectively.} \label{fig3:Overplots}
\end{figure}

The filling strategy shows obvious success in the adjustment of the scaling indices map (see the lower right panel of figure \ref{fig1:dreimaldrei}): The white noise leads to higher values in the $\alpha$-response for the pixels close to the mask as compared to the masking method. The regions around the edges of the mask feature now $\alpha$-responses that match far better the values of the remaining regions. Still these $\alpha$-values are calculated with the help of an artificial environment, but now the contortions are lower compared to the original approach. Since we apply this method to both the original WMAP data as well as to the simulations, the now smaller systematic errors in the $\alpha$-calculation for the 'edge'-pixels are the same for both kinds of maps. Thus any significant deviation found in the WMAP data is due to intrinsic effects. \par

\begin{figure}
\centering
\includegraphics[width=8cm, keepaspectratio=true]{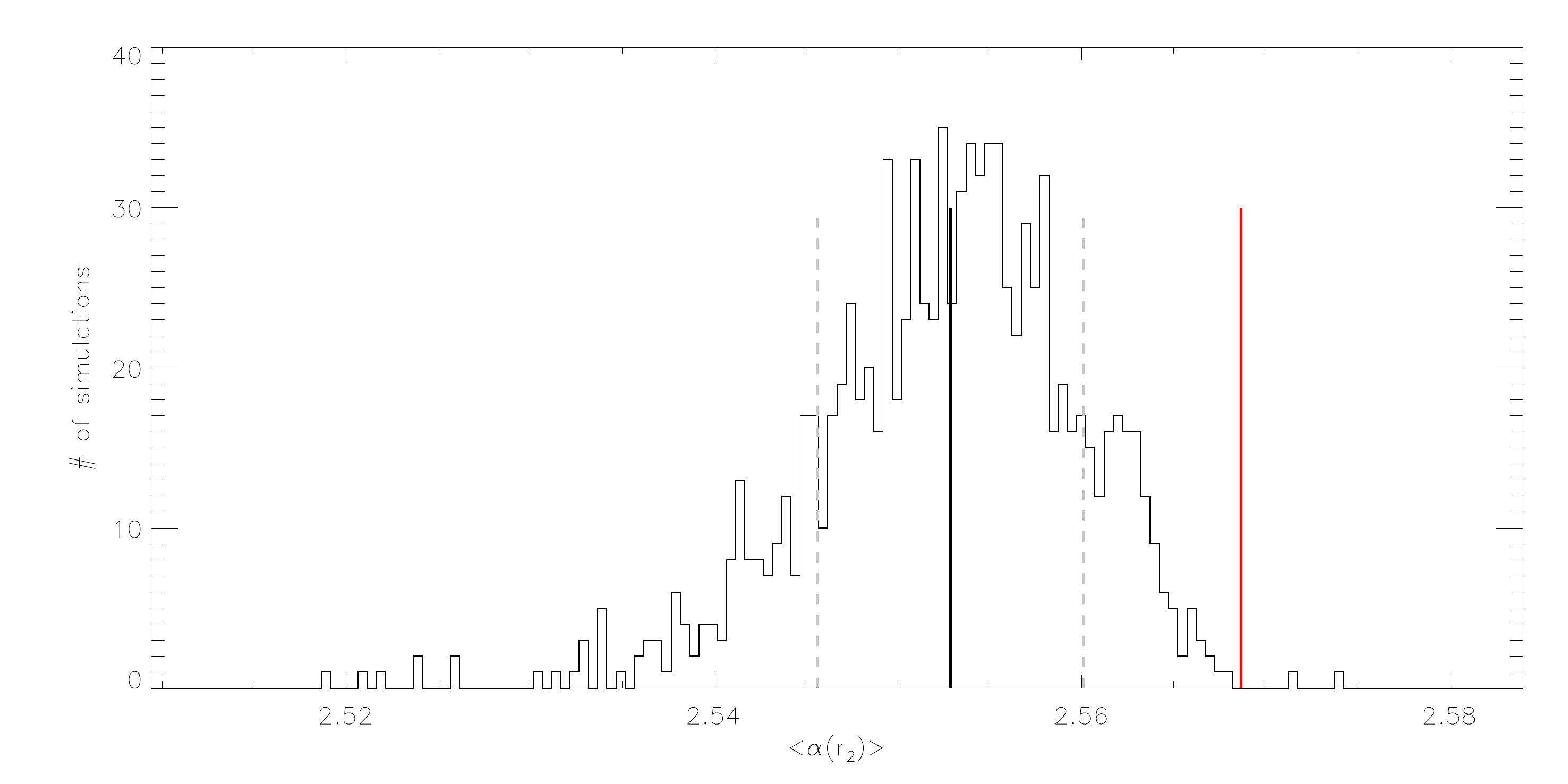}
\includegraphics[width=8cm, keepaspectratio=true]{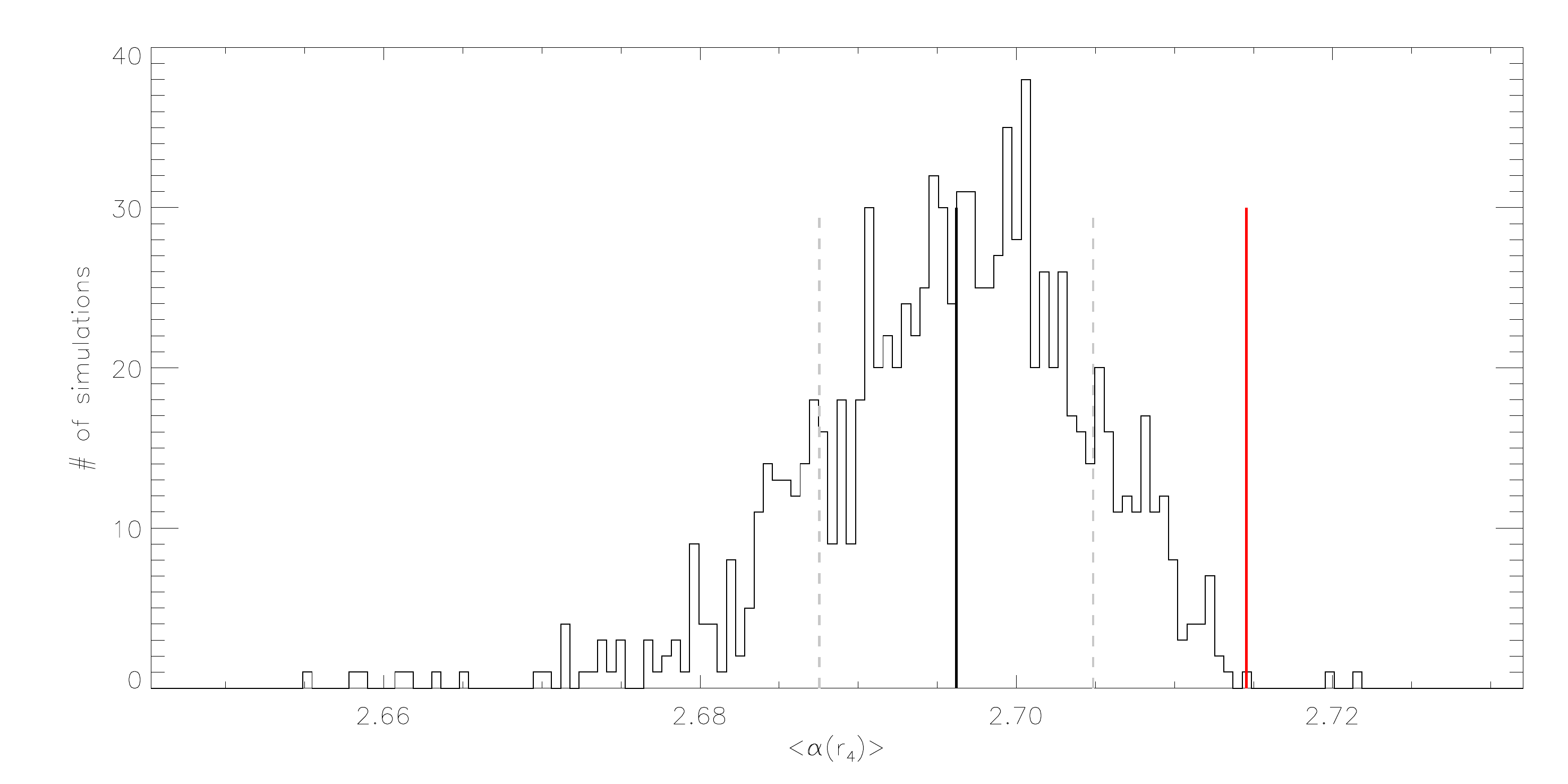}
\includegraphics[width=8cm, keepaspectratio=true]{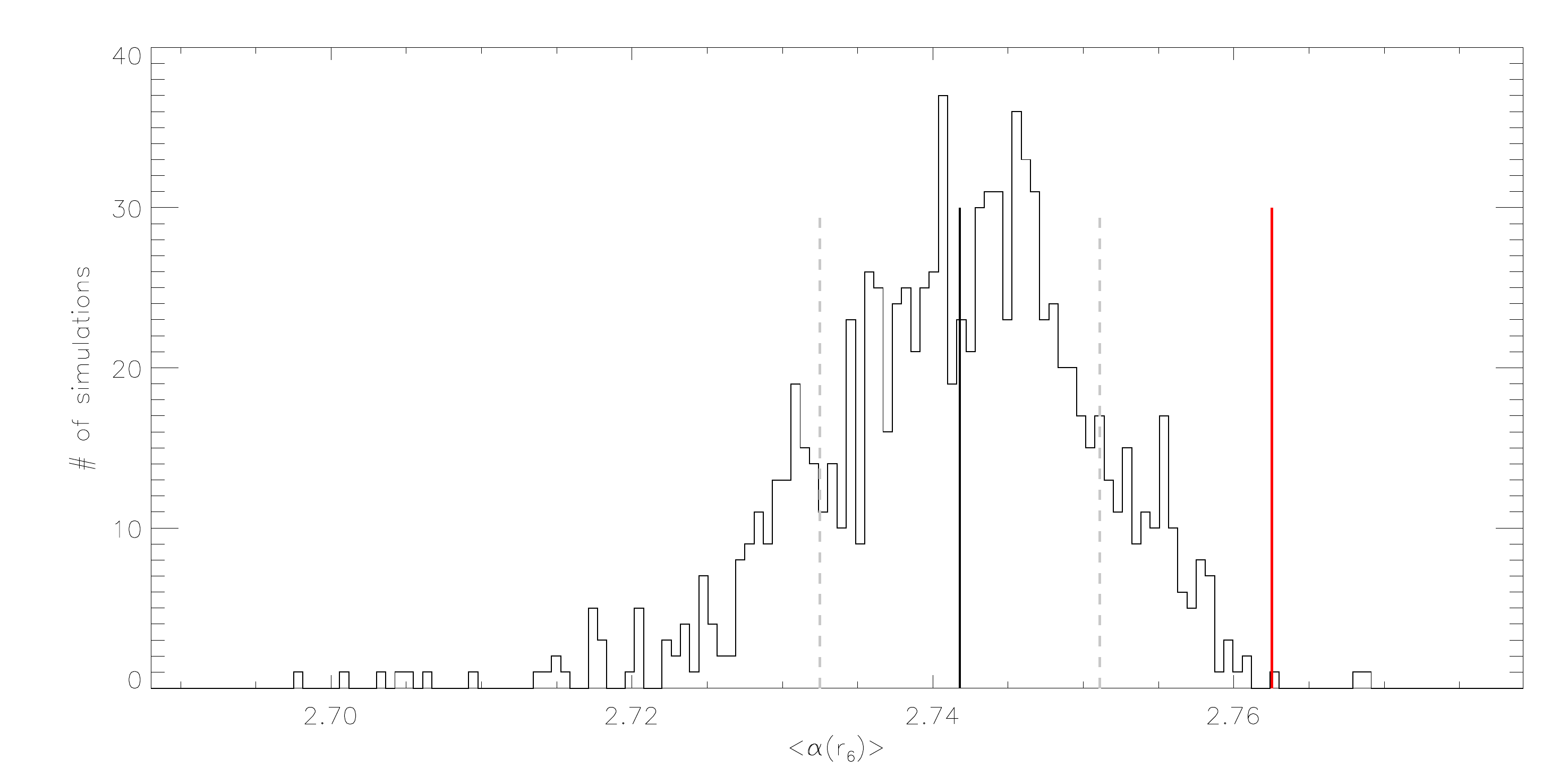}
\includegraphics[width=8cm, keepaspectratio=true]{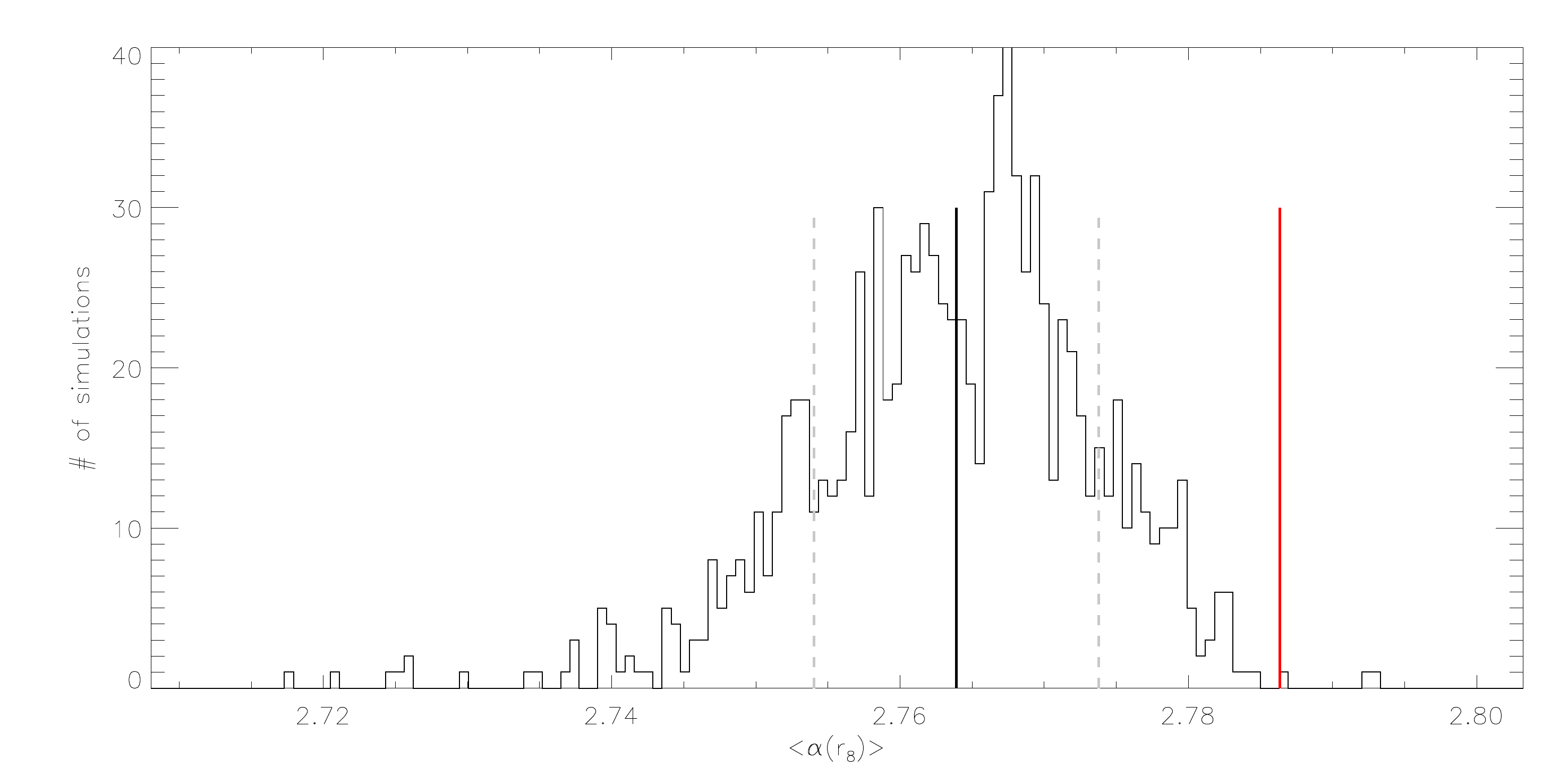}
\includegraphics[width=8cm, keepaspectratio=true]{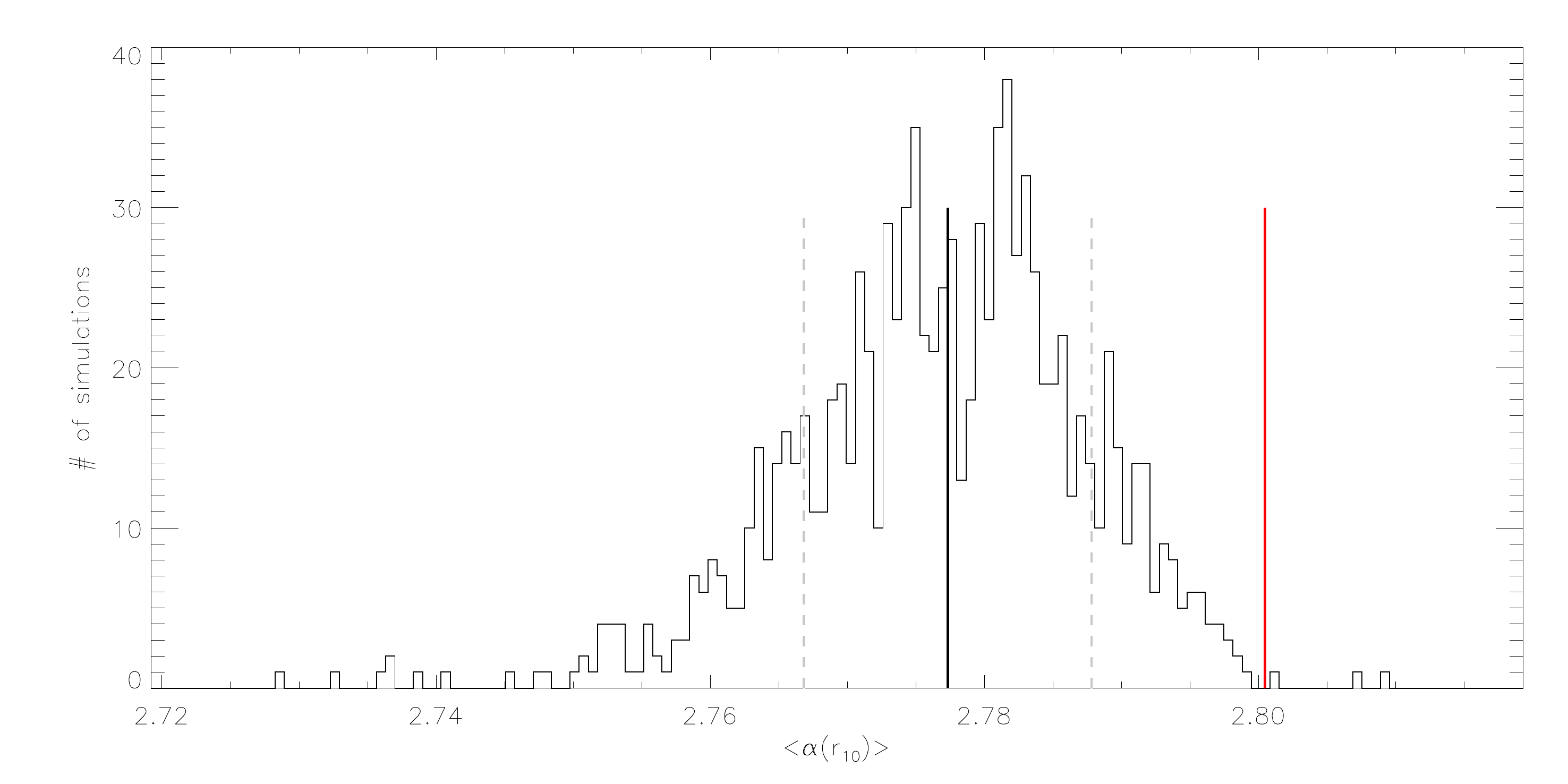}
\caption{The histograms of the mean values $\langle \alpha(r_k) \rangle$ of the 1000 simulations for the five different scale parameters $r_2$, $r_4$, $r_6$, $r_8$ and $r_{10}$, calculated for the full sky. The red lines denote the corresponding results of the WMAP data, while the black and grey lines characterise the average over all simulations and the $1\sigma$ regions, respectively.} \label{fig3.5:MeanComparison}
\end{figure}

The most important advantage of the filling strategy arises if one considers local features in the CMB map: For a search of local anomalies (e.g. cold spots), the filled map provides a better underlying than the original map: In a point distribution, spots as well as points at the border of the distribution show similarly low $\alpha$-values. Thus, by cutting out the mask, it is difficult to decide whether a detected local feature really exists or whether it originates from a masked area nearby. But by using the filling technique, there is no longer an edge between the masked and the non-masked regions, and anomalies in the Gaussian noise of the mask are highly improbable. Thus, any detected local feature must then originate from the data itself. Considering the amount of masked areas \textit{outside} the galactic plane, this technique describes a considerable improvement for the whole sky. Due to these advantages we will only use the mask-filled maps in chapter \ref{localfeatures}.


\section{Results}

\subsection{Band-wise and co-added map analysis}
\label{VierEins}

In figure \ref{fig3:Overplots}, the empirical probability densities $P(\alpha)$ of the scaling indices (calculated with $r=0.2$) are displayed for the WMAP data and for the simulations, evaluated for the original and the filling method from chapter \ref{Copingboundary}. For clarity reasons we only used the first 50 simulations in these plots. For both methods, a shift of the WMAP data to higher values can be detected, that becomes particularly apparent in the northern hemishpere of the galactic coordinate system. This indicates a more 'unstructured' arrangement of the underlying temperature fluctuations of the CMB data in comparison to the simulations. In addition, the histograms of the simulations are slightly broader and therewith containing a larger structural variability than the one of the WMAP data. \par
Comparing the non-filling and the filling method, the histograms of the latter feature a higher maximum as well as higher values for large $\alpha$, but lower probabilities for $\alpha \in [2.0,2.5]$. The obvious reason for this shift is the fact that the filled mask does not reduce the $\alpha$-values of its surroundings as it was the case with the former method. Now, the outcome of these regions is influenced by the white noise and is therefore allocated at higher values. \par
If we focus on the mean values $\langle \alpha(r_k) \rangle$ of the scaling indices, and compare the results of the simulations with the WMAP data, the above mentioned shift to higher values becomes yet clearer. This can be seen in Figure \ref{fig3.5:MeanComparison}, where the distribution of $\langle \alpha(r_k) \rangle$ for the simulations as well as the data is displayed for the five different scale parameters $r_2$, $r_4$, $r_6$, $r_8$ and $r_{10}$. These results were obtained using the full sky. For all applied scales, the distance between the average over all simulations and the result of the original WMAP map is notably similar. If we perform this analysis for the northern hemisphere only, the deviations of the original data as compared to the simulations become significantly larger. \par

\begin{figure}
\centering
\includegraphics[width=8cm, keepaspectratio=true]{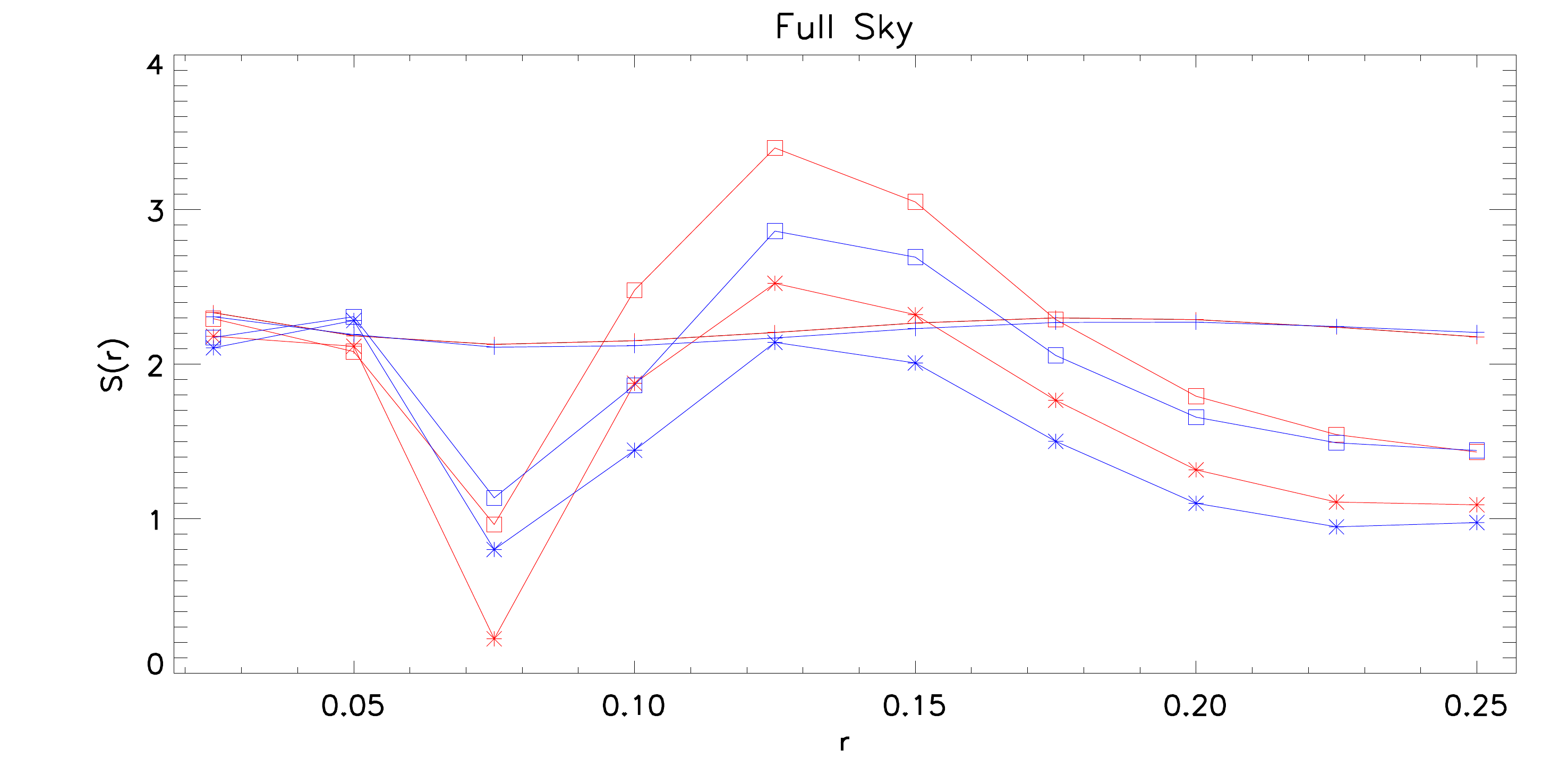}
\includegraphics[width=8cm, keepaspectratio=true]{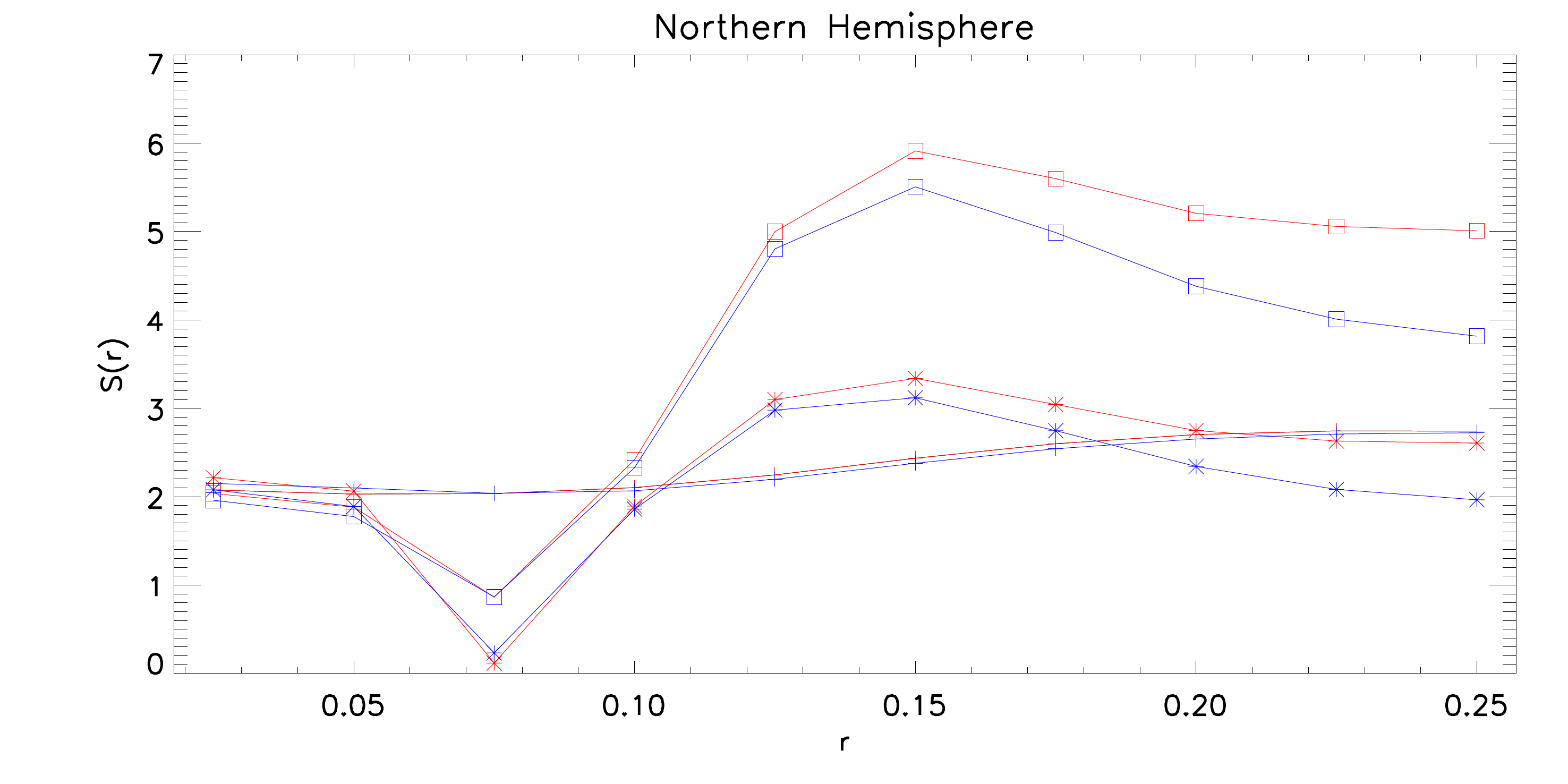}
\includegraphics[width=8cm, keepaspectratio=true]{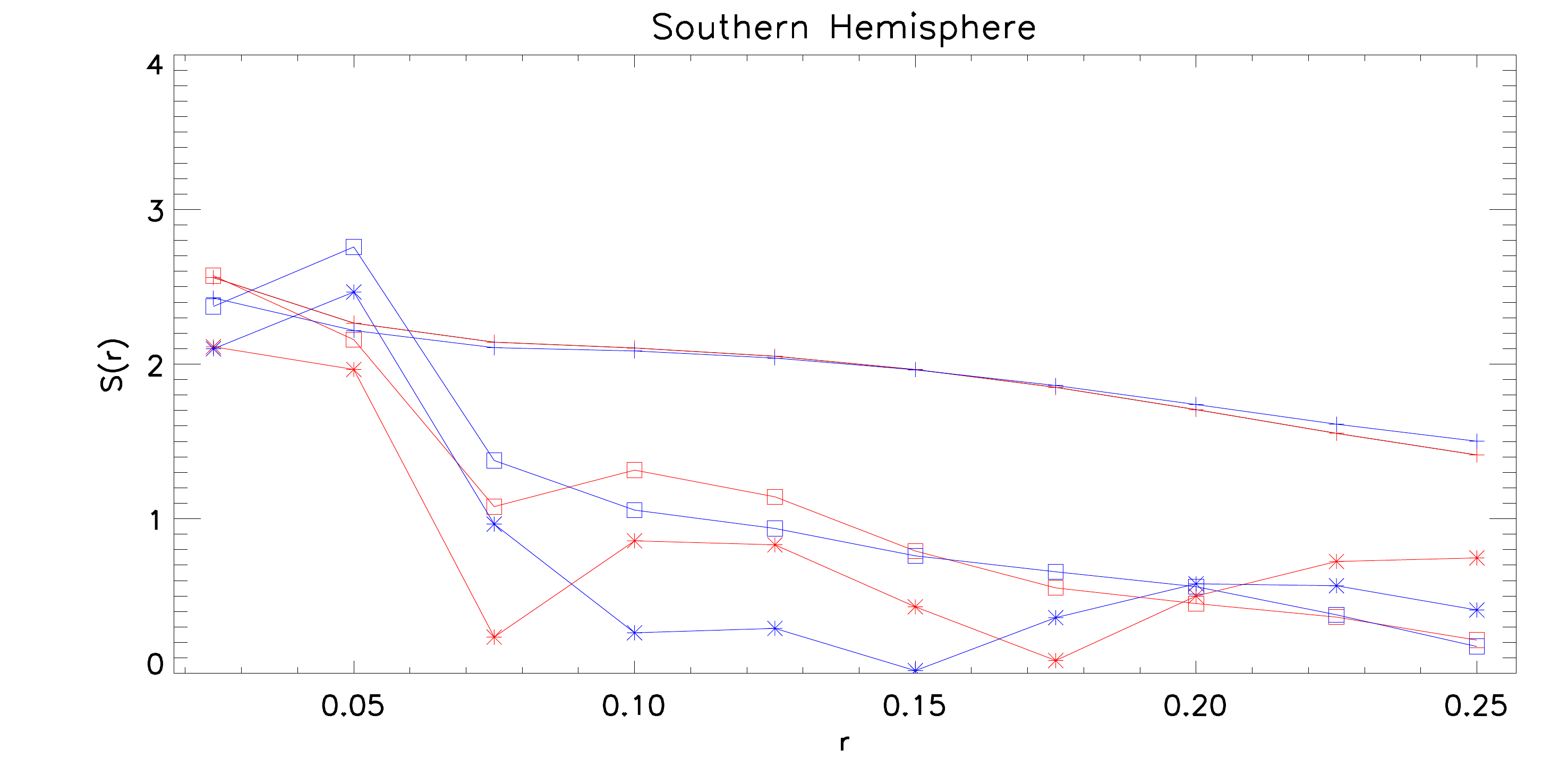}
\caption{The $\sigma$-normalised deviations $S(r)$ of the statistics of the equations (\ref{MeanStdevChi}) - (\ref{Zeile3}) in absolute values for the VW-band, plotted as a function of the scale parameter $r$. The lines with "$+$" denote the mean, "$*$" the standard deviation and the boxes the $\chi^2$-combination, each for the original (red) and the mask filling method (blue). As in figure \ref{fig3:Overplots}, the upper diagram shows the results of a full-sky analysis, while the middle and the lower ones show the results when only concerning the northern or southern hemisphere respectively.} \label{fig4:SigVW}
\end{figure}

\begin{figure}
\centering
\includegraphics[width=8cm, keepaspectratio=true]{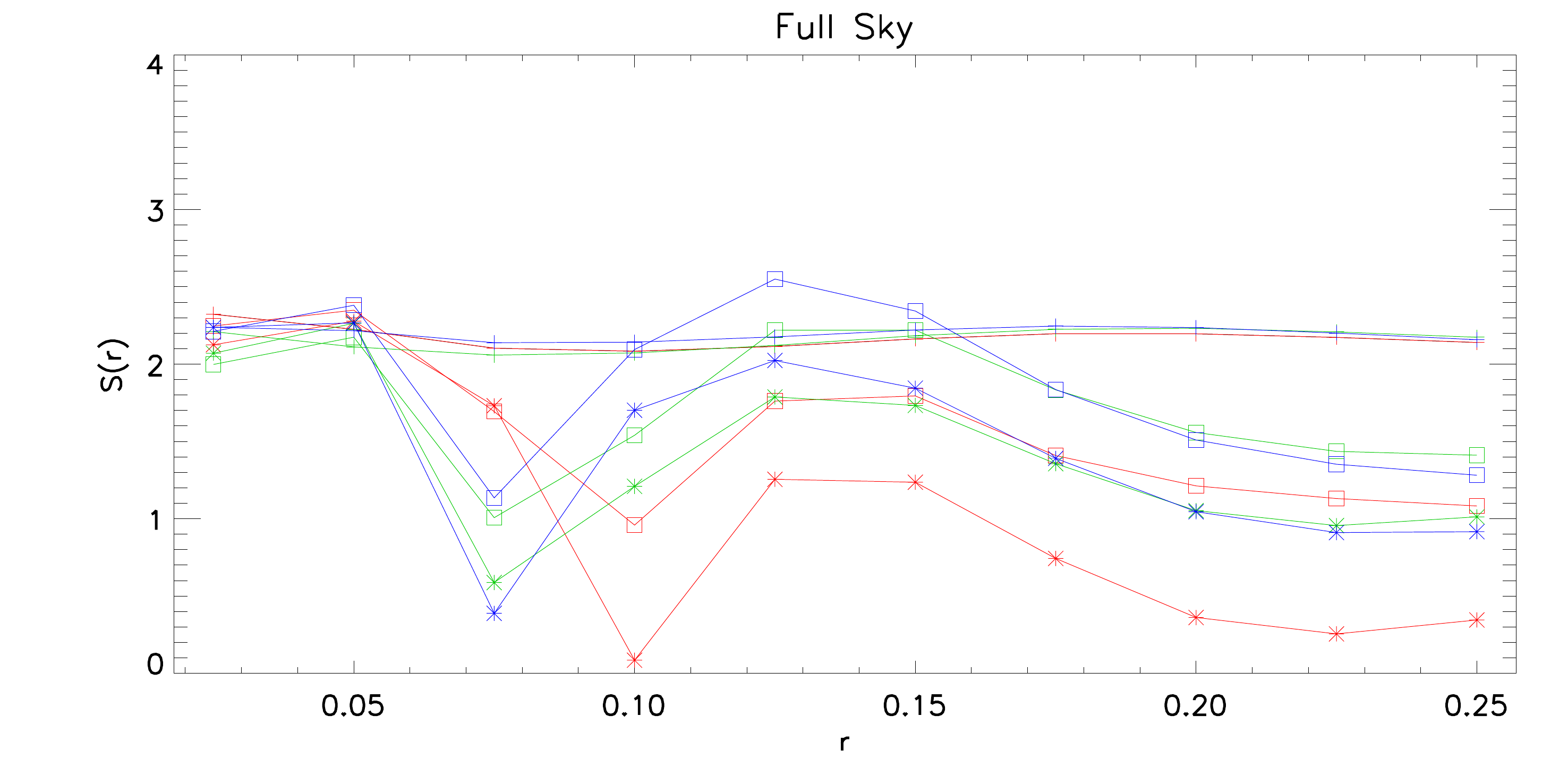}
\includegraphics[width=8cm, keepaspectratio=true]{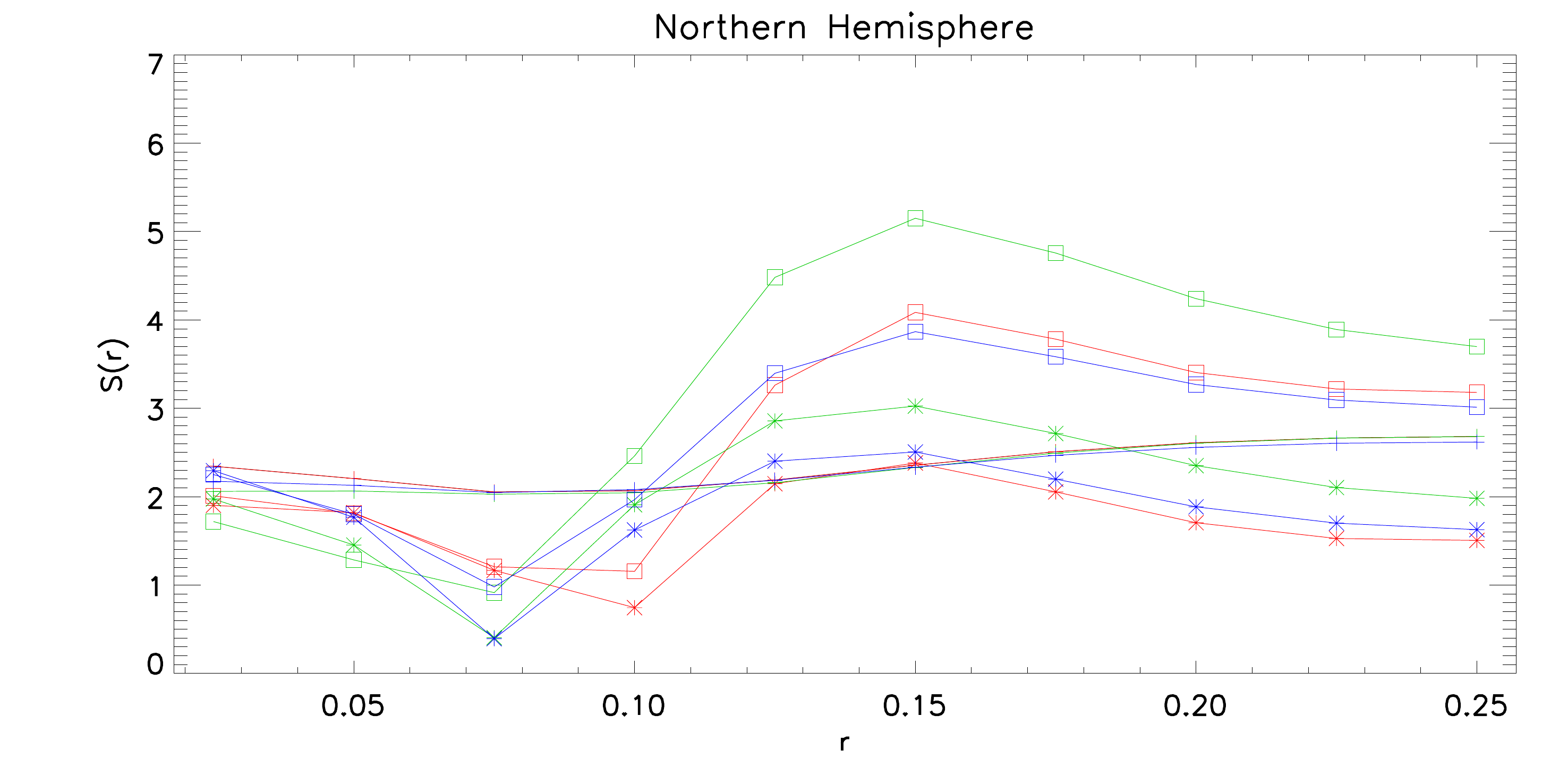}
\includegraphics[width=8cm, keepaspectratio=true]{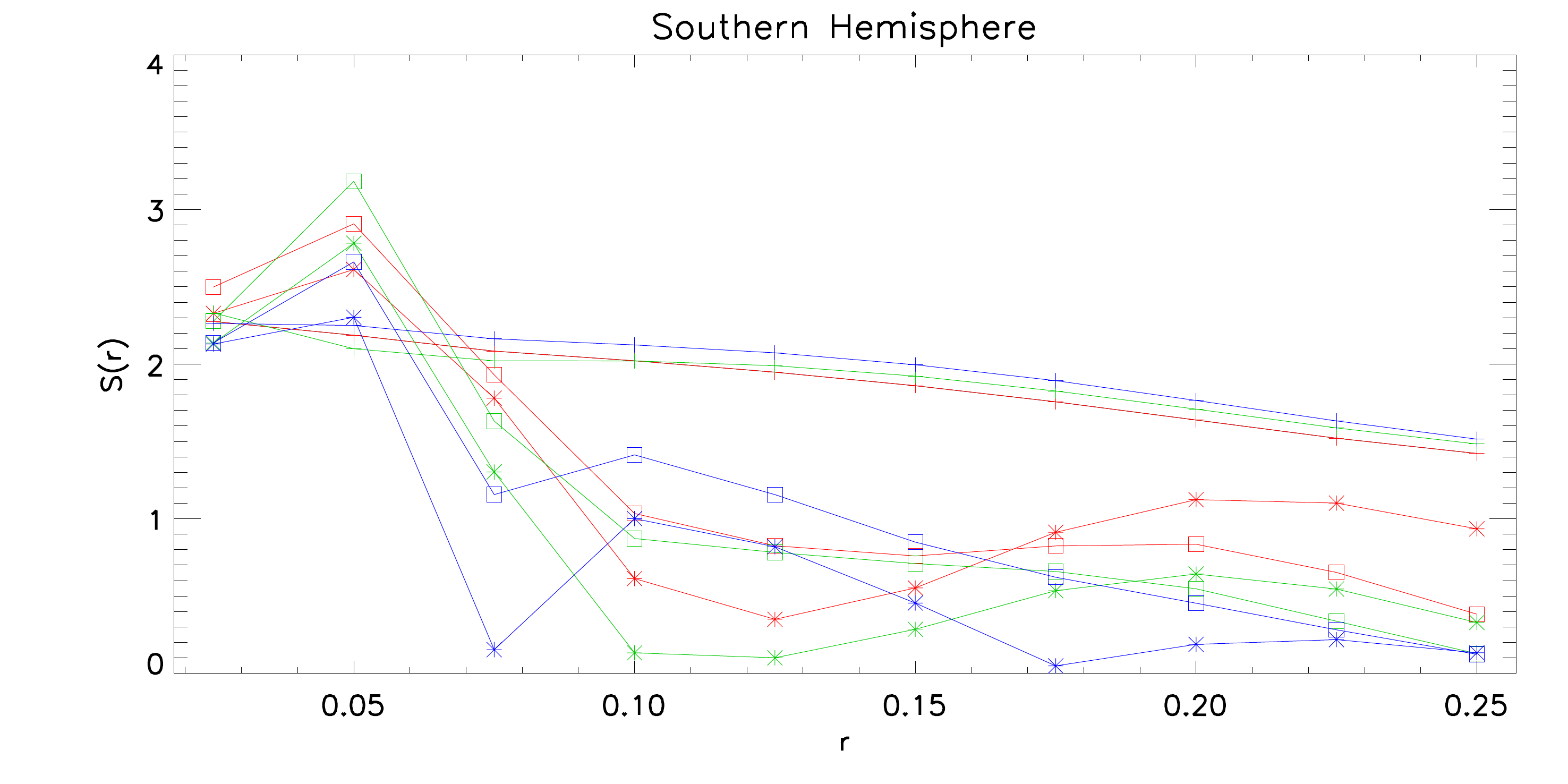}
\caption{Same as figure \ref{fig4:SigVW} but applied to the Q- (red), V- (green) and W-band (blue). Only the results of the mask-filling method are shown while the original method is left out.} \label{fig5:SigQVW}
\end{figure}

Both the shift to higher values of the WMAP data in comparison to the simulations as well as its broader density are reflected in the $\sigma$-normalised deviations $S(r)$ of the scale-dependent statistics of the equations (\ref{MeanStdevChi}) - (\ref{Zeile3}): The former is reflected in the mean and the latter in the standard deviation (and therefore both aspects in the diagonal $\chi^2$-statistics). Figure \ref{fig4:SigVW} shows these deviations for the coadded VW-band as a function of the scale parameter $r$ for the original and the mask filling method, while figure \ref{fig5:SigQVW} displays only the latter method, but for the three single bands. As above, the results are illustrated for the full sky as well as for the seperate hemispheres. The shift to higher values of the WMAP data in the northern hemisphere in figure \ref{fig3:Overplots} appears now as an increased $S(r)$, especially for higher scales ($r \geq 0.125$), where the deviations of the two moments range between $2\sigma$ and $3.5\sigma$, and the $\chi^2$-combination nearly reaches a $6\sigma$-level. In the southern hemisphere, only the lowest scales show a namable $S(r)$. On larger scales, no signatures for deviations from Gaussianity are identified. Looking at the single bands Q, V and W, the overall qualitative behaviour of the images is quite similar, while the $\sigma$-normalised deviations itself are slightly lower in most cases. A remarkable fact is the appearance of the \textit{highest} $S(r)$ ($5.2\sigma$ for the $\chi^2$-combination in the northern hemisphere at $r=0.15$) in the frequency band which is considered to be the \textit{least} foreground-contaminated one, namely the V-band. Comparing the co-added VW-band of the original approach and of the mask filling method, the $\sigma$-normalised deviations of the mean are almost identical. The standard deviation of the latter method in comparison to the former one shows a slightly lower $S(r)$ for higher scales, which is also reflected in the graph of the $\chi^2$-combination, yet the profile remains the same. \par

\begin{table}
\begin{tabular}{lcccc}
\hline \hline
 &  Full Sky  &  Northern Sky & Southern Sky \\ \hline
$\chi^2_{\langle \alpha \rangle}$: & $(S/\%)$ & $(S/\%)$ & $(S/\%)$ \\ \\
VW (original)       & 2.2 / 97.3 & 2.7 / 98.5 & 1.7 / 96.6 \\
VW (mask-filled) & 2.2 / 97.3 & 2.7 / 98.3 & 1.7 / 96.6 \\
Q    (mask-filled) & 2.1 / 97.4 & 2.7 / 98.3 & 1.5 / 95.8 \\
V     (mask-filled) & 2.0 / 97.4 & 2.6 / 98.1 & 1.5 / 96.2 \\
W    (mask-filled) & 2.1 / 97.5 & 2.6 / 98.2 & 1.7 / 96.6 \\ \hline
$\chi^2_{\sigma_{\alpha}}$: & & & \\ \\
VW (original)       & 2.0 / 95.6 & 5.5 / 99.7 & 0.1 / 68.5 \\
VW (mask-filled) & 1.6 / 93.4 & 4.3 / 99.3 & 0.3 / 72.9 \\
Q    (mask-filled) & 0.7 / 83.1 & 2.3 / 96.0 & 1.2 / 89.4 \\
V    (mask-filled) & 1.2 / 90.5 & 4.0 / 98.9 & 0.6 / 79.2 \\
W   (mask-filled) & 1.4 / 92.3 & 2.8 / 96.7 & 1.9 / 71.6 \\ \hline
$\chi^2_{\langle \alpha \rangle,\sigma_{\alpha}}$: & & & \\ \\
VW (original)       & 2.3 / 97.4 & 4.2 / 99.1 & 1.3 / 93.5 \\
VW (mask-filled) & 2.1 / 97.1 & 3.7 / 98.8 & 1.3 / 94.2 \\
Q    (mask-filled) & 1.8 / 96.3 & 2.9 / 98.3 & 1.6 / 95.5 \\
V    (mask-filled) & 1.9 / 96.6 & 3.5 / 98.8 & 1.3 / 94.1 \\
W   (mask-filled) & 2.0 / 96.4 & 3.0 / 98.5 & 1.3 / 93.7 \\ \hline \hline
\end{tabular}
\caption{The $\sigma$-normalised deviations $S$ and the empirical probabilities $p$ of the scale-independent diagonal $\chi^2$-statistics from the equations (\ref{ChiSummed}) to (\ref{ZeileC}) for the different bands and methods as well as for the Full Sky and the single hemispheres.} \label{Table1}
\end{table}

\begin{table}
\begin{tabular}{lcccc}
\hline \hline
 &  Full Sky  &  Northern Sky & Southern Sky \\ \hline
$\langle \alpha (0.2) \rangle$: & $(S/\%)$ & $(S/\%)$ & $(S/\%)$ \\ \\
VW (original)       & 2.3 / 99.7 & 2.7 / 99.8 & 1.7 / 97.1 \\
VW (mask-filled) & 2.3 / 99.7 & 2.3 / 98.7 & 1.7 / 97.6 \\
Q    (mask-filled) & 2.2 / 99.6 & 2.6 / 99.8 & 1.6 / 96.6 \\
V     (mask-filled) & 2.2 / 99.6 & 2.6 / 99.8 & 1.7 / 97.2 \\
W    (mask-filled) & 2.2 / 99.6 & 2.6 / 99.8 & 1.8 / 98.2 \\ \hline
$\sigma_{\alpha(0.2)}$: & & & \\ \\
VW (original)       & 1.3 / 90.7 & 2.7 / 99.8 & 0.5 / 68.9 \\
VW (mask-filled) & 1.1 / 85.6 & 4.3 / 99.3 & 0.6 / 69.8 \\
Q    (mask-filled) & 0.4 / 64.2 & 1.7 / 95.6 & 1.1 / 85.8 \\
V    (mask-filled) & 1.1 / 84.9 & 2.4 / 99.0 & 0.6 / 71.4 \\
W   (mask-filled) & 1.0 / 83.9 & 1.9 / 96.4 & 0.2 / 55.9 \\ \hline
$ \chi^2_{\alpha (0.2)}$: & & & \\ \\
VW (original)       & 1.8 / 95.3 & 5.2 / 99.4 & 0.5 / 81.6 \\
VW (mask-filled) & 1.7 / 94.5 & 4.4 / 99.1 & 0.6 / 82.6 \\
Q    (mask-filled) & 1.2 / 90.9 & 3.4 / 98.6 & 0.8 / 97.1 \\
V    (mask-filled) & 1.6 / 94.2 & 4.2 / 99.2 & 0.5 / 81.9 \\
W   (mask-filled) & 1.5 / 93.7 & 3.3 / 98.6 & 0.5 / 81.0 \\ \hline \hline
\end{tabular}
\caption{Same as table \ref{Table1}, but for the scale-dependent statistics from the equations (\ref{MeanStdevChi}) to (\ref{Zeile3}) for the single scale $r=0.2$.} \label{Table1.5}
\end{table}

We also calculated the $\sigma$-normalised deviations $S$ and the percentages $p$ of the simulations with higher (lower) results of the scale-independent diagonal $\chi^2$-statistics from the equations (\ref{ChiSummed}) to (\ref{ZeileC}), which are listed in table \ref{Table1}. Although the results are damped by a few unimportant scales, high deviations are still found, particularly in the northern hemisphere. For a better comparison to seperate scale lengths, the respective results of the scale-dependent statistics (\ref{MeanStdevChi}) to (\ref{Zeile3}) are listed in table \ref{Table1.5}, for which we used the single scale $r=0.2$. \par
In general, all occuring characteristics of the figures \ref{fig3:Overplots} and \ref{fig4:SigVW} match the findings of the analysis of the WMAP 3-year data in \citet{raeth07a}. This indicates that the results are not based on some time-dependent effects. Since the 5-year data features lower error bars than the 3-year data, it is also improbable that both results are induced by noise effects only. \par

\begin{figure*}
\centering
\includegraphics[width=5.8cm, keepaspectratio=true]{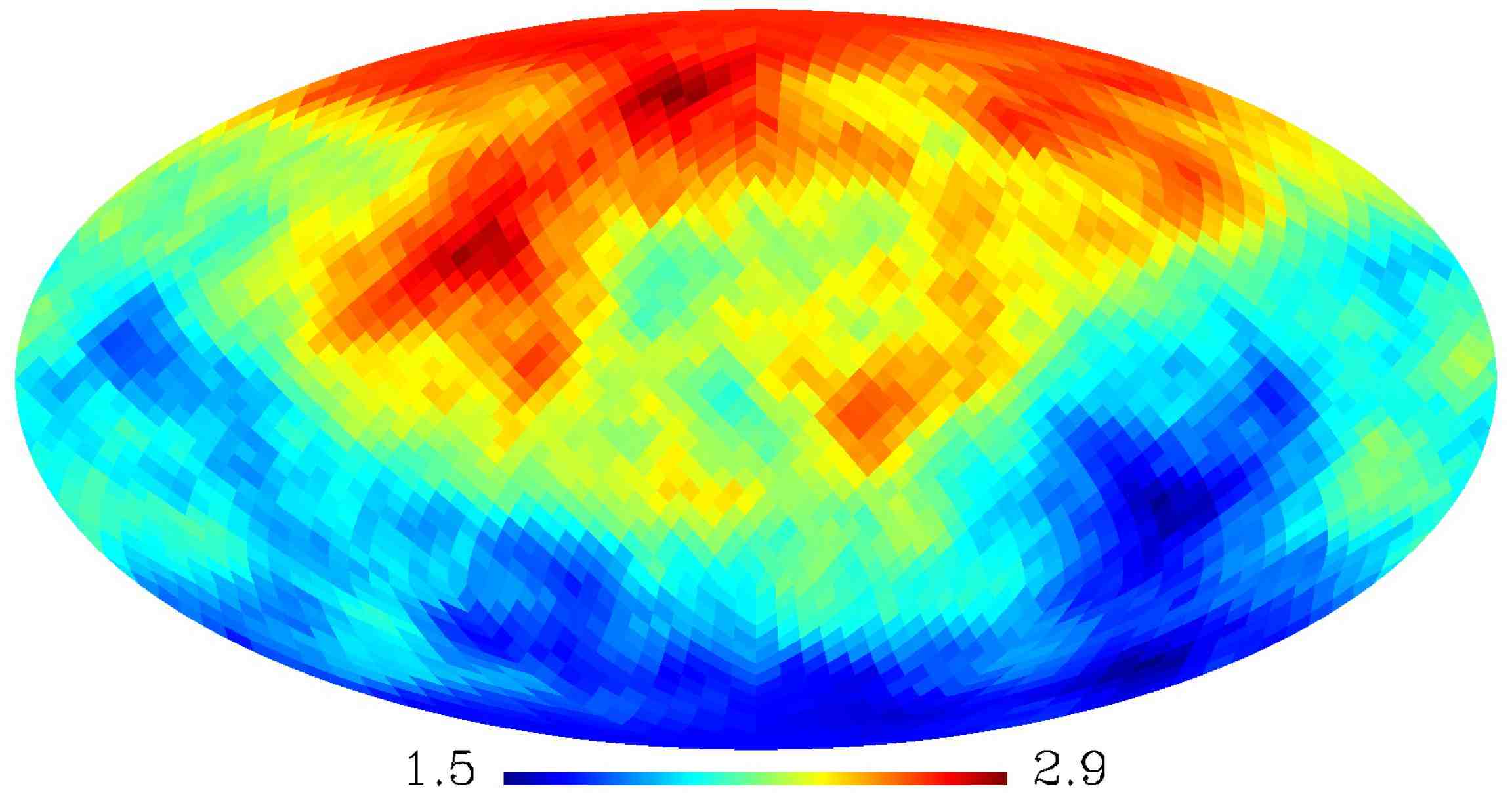}
\includegraphics[width=5.8cm, keepaspectratio=true]{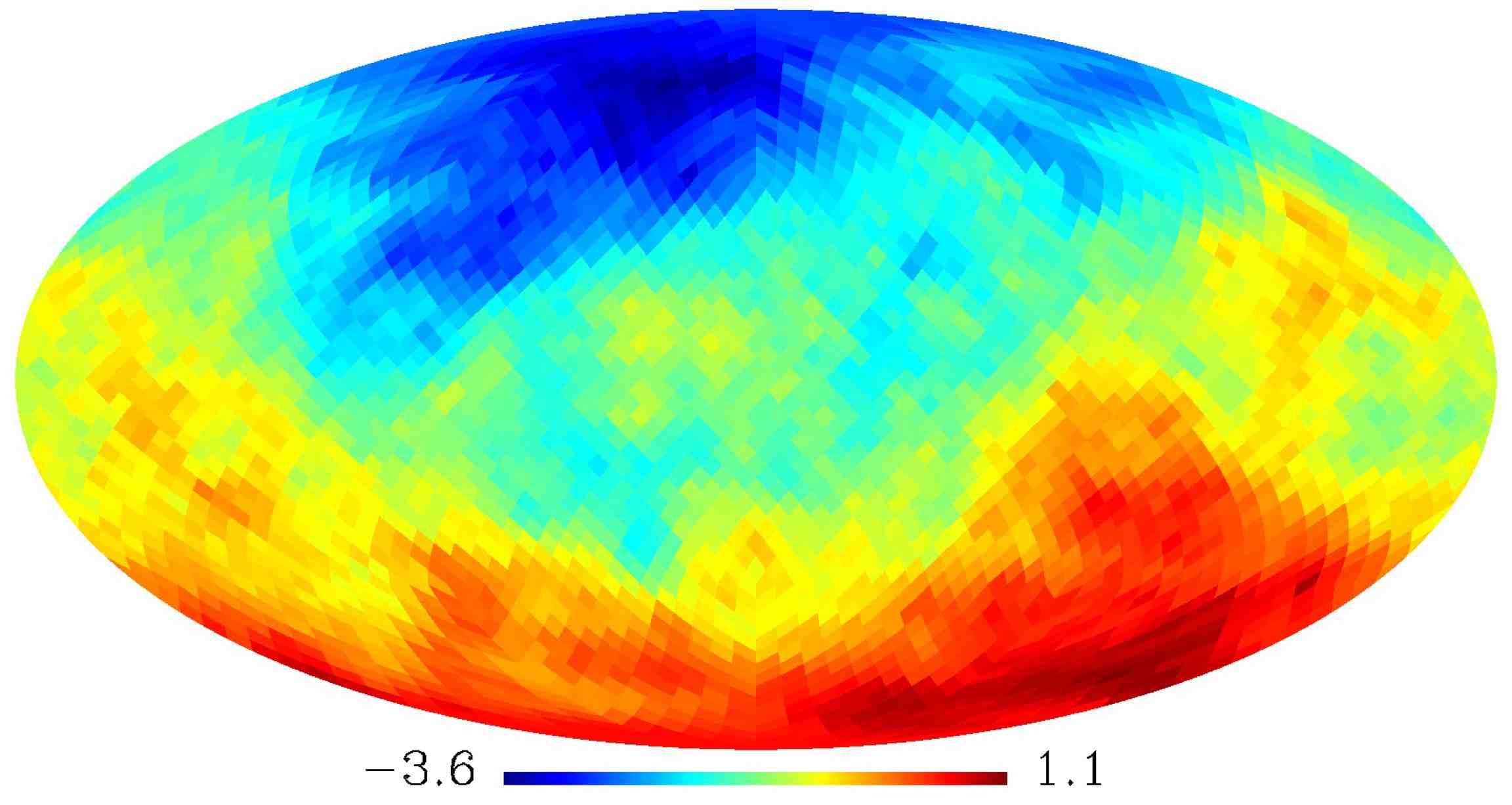}
\includegraphics[width=5.8cm, keepaspectratio=true]{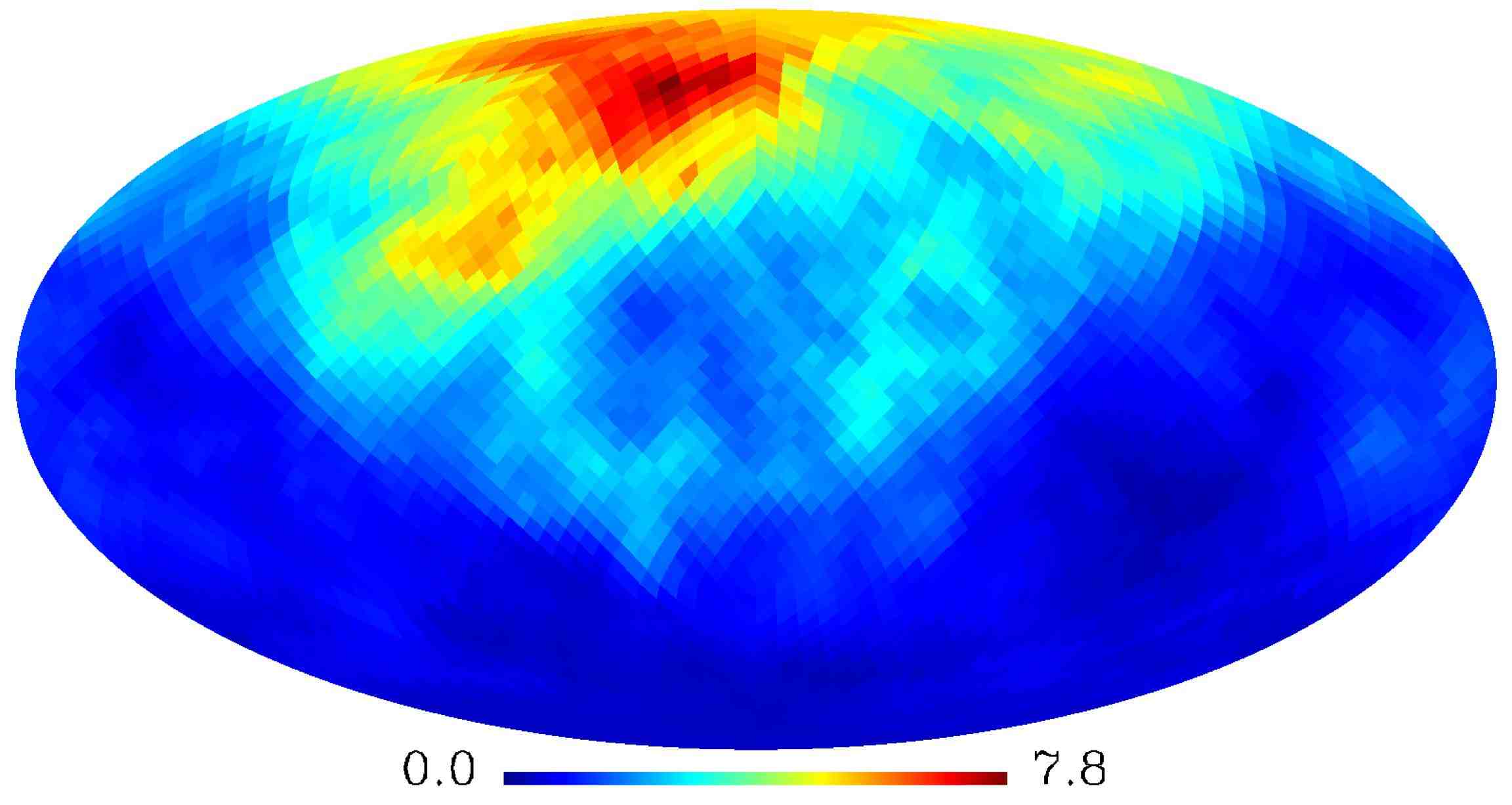}
\vspace{9mm}
\includegraphics[width=5.8cm, keepaspectratio=true]{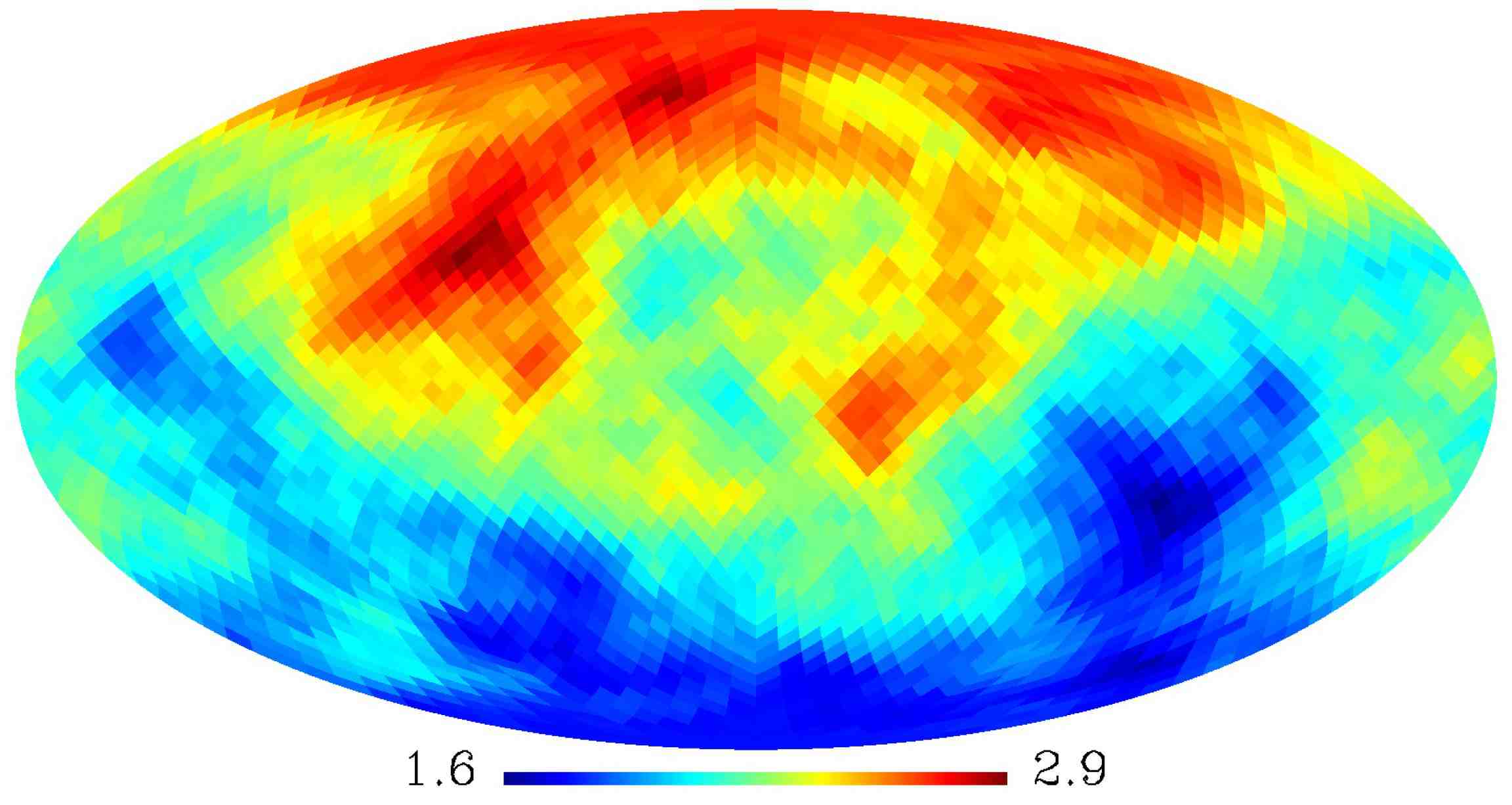}
\includegraphics[width=5.8cm, keepaspectratio=true]{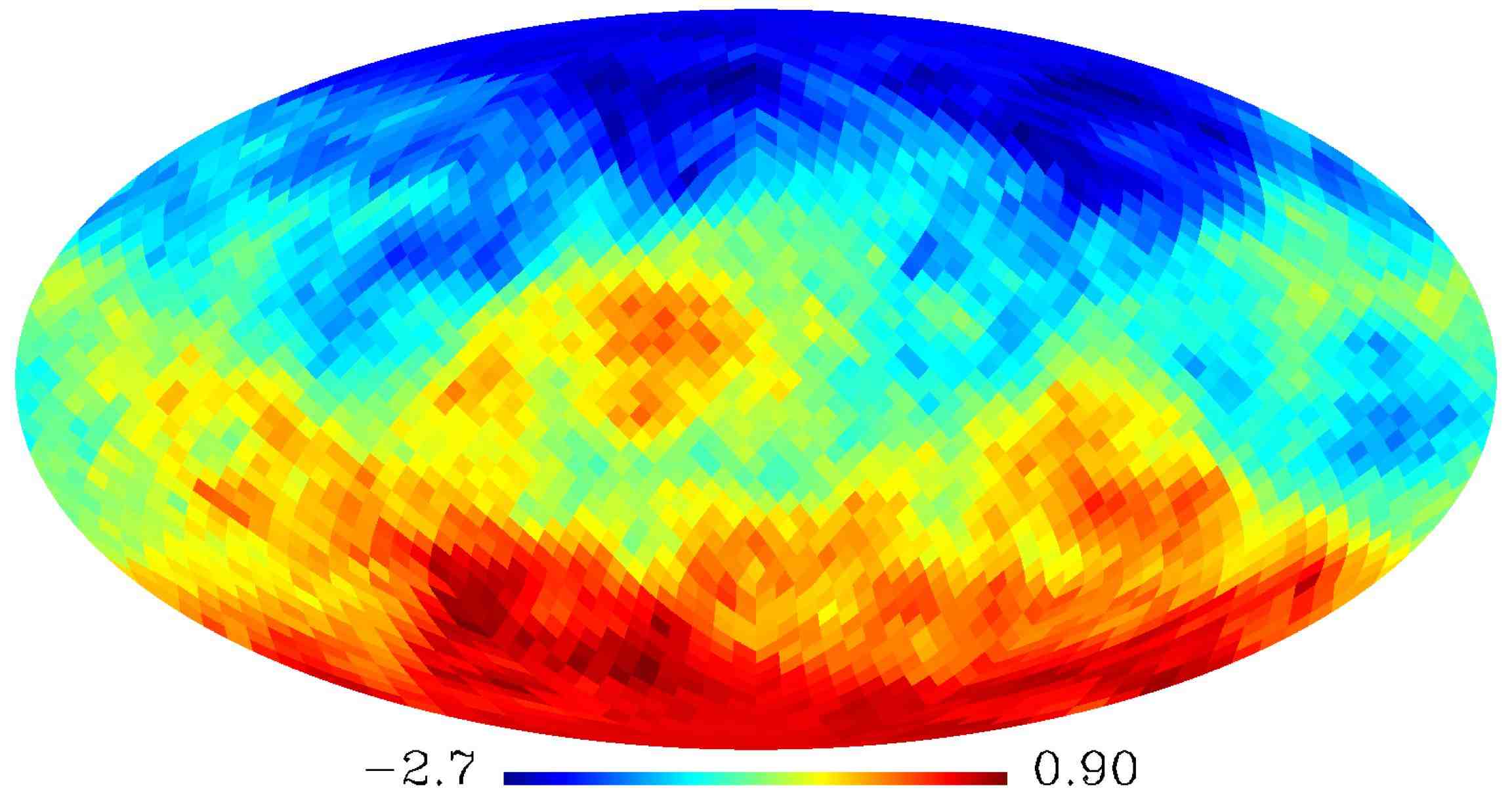}
\includegraphics[width=5.8cm, keepaspectratio=true]{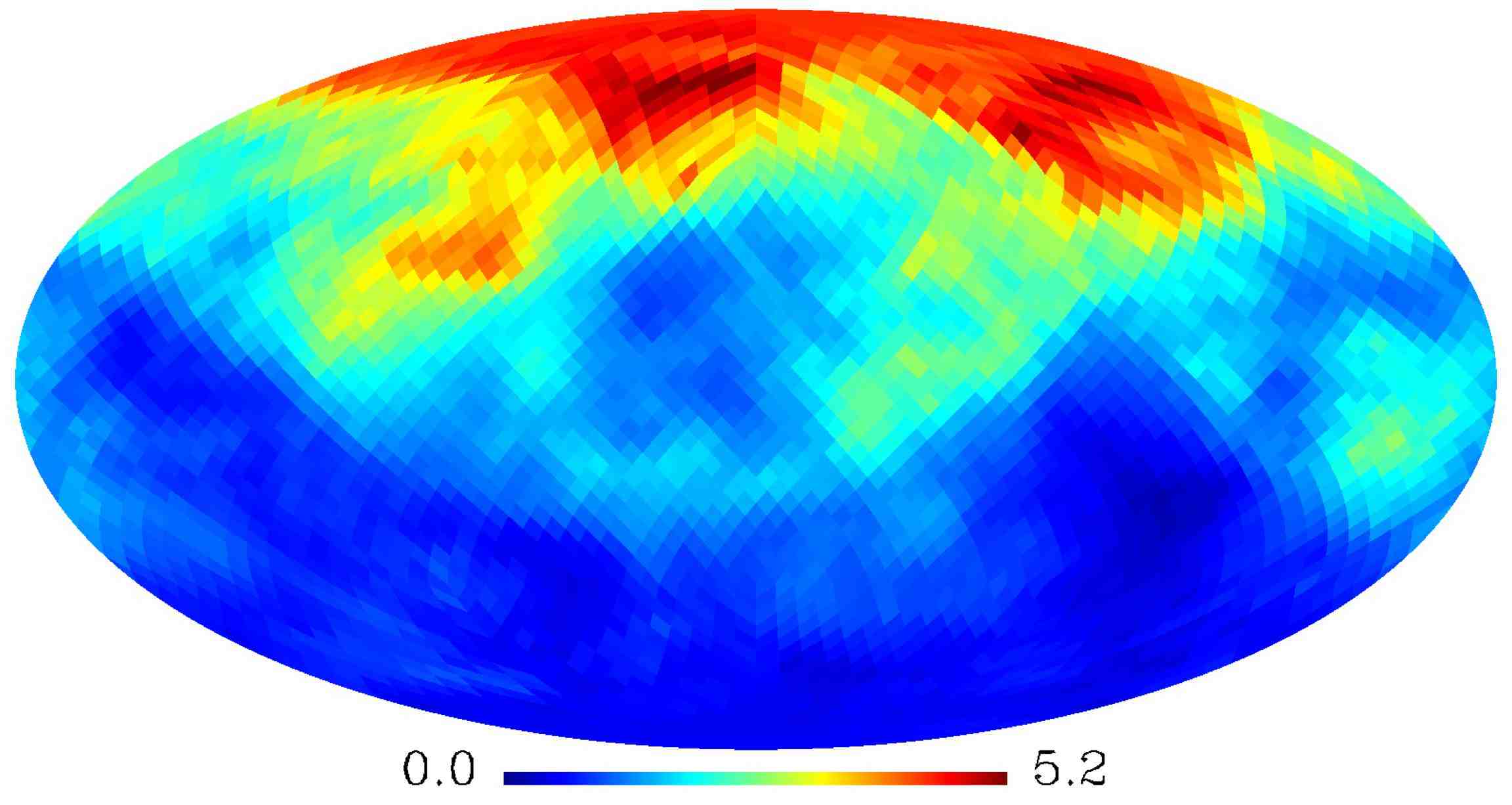}
\includegraphics[width=5.8cm, keepaspectratio=true]{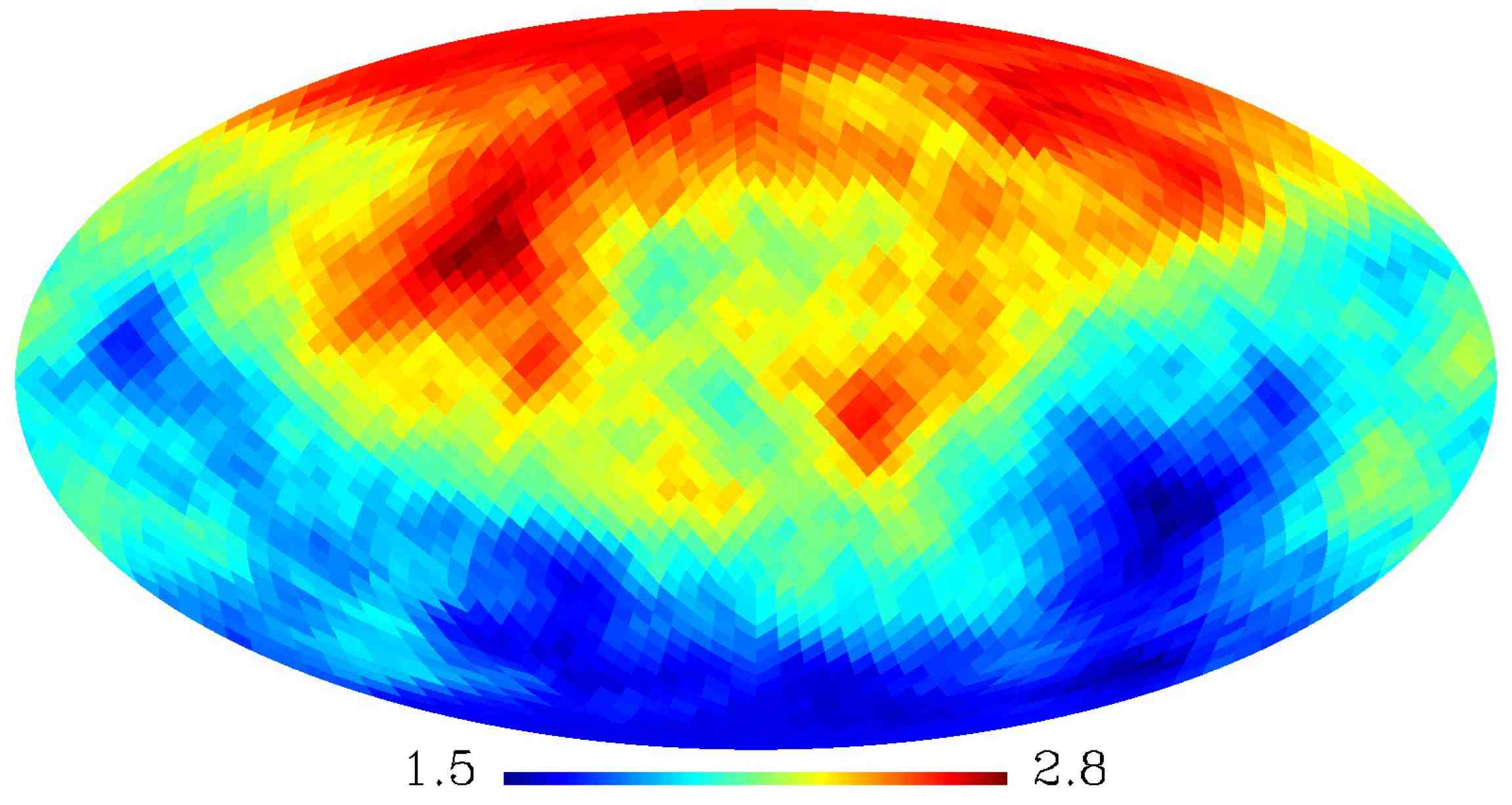}
\includegraphics[width=5.8cm, keepaspectratio=true]{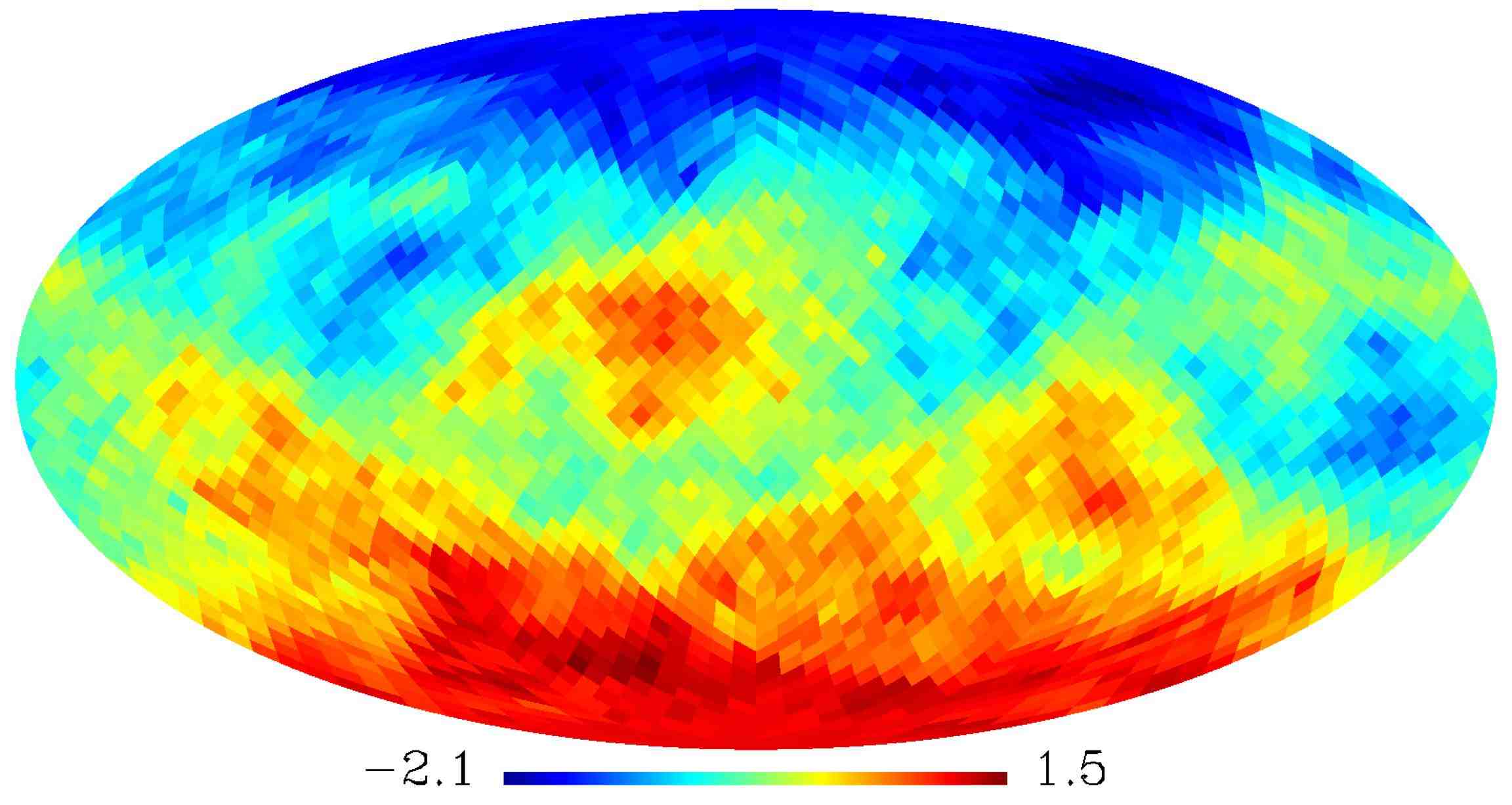}
\includegraphics[width=5.8cm, keepaspectratio=true]{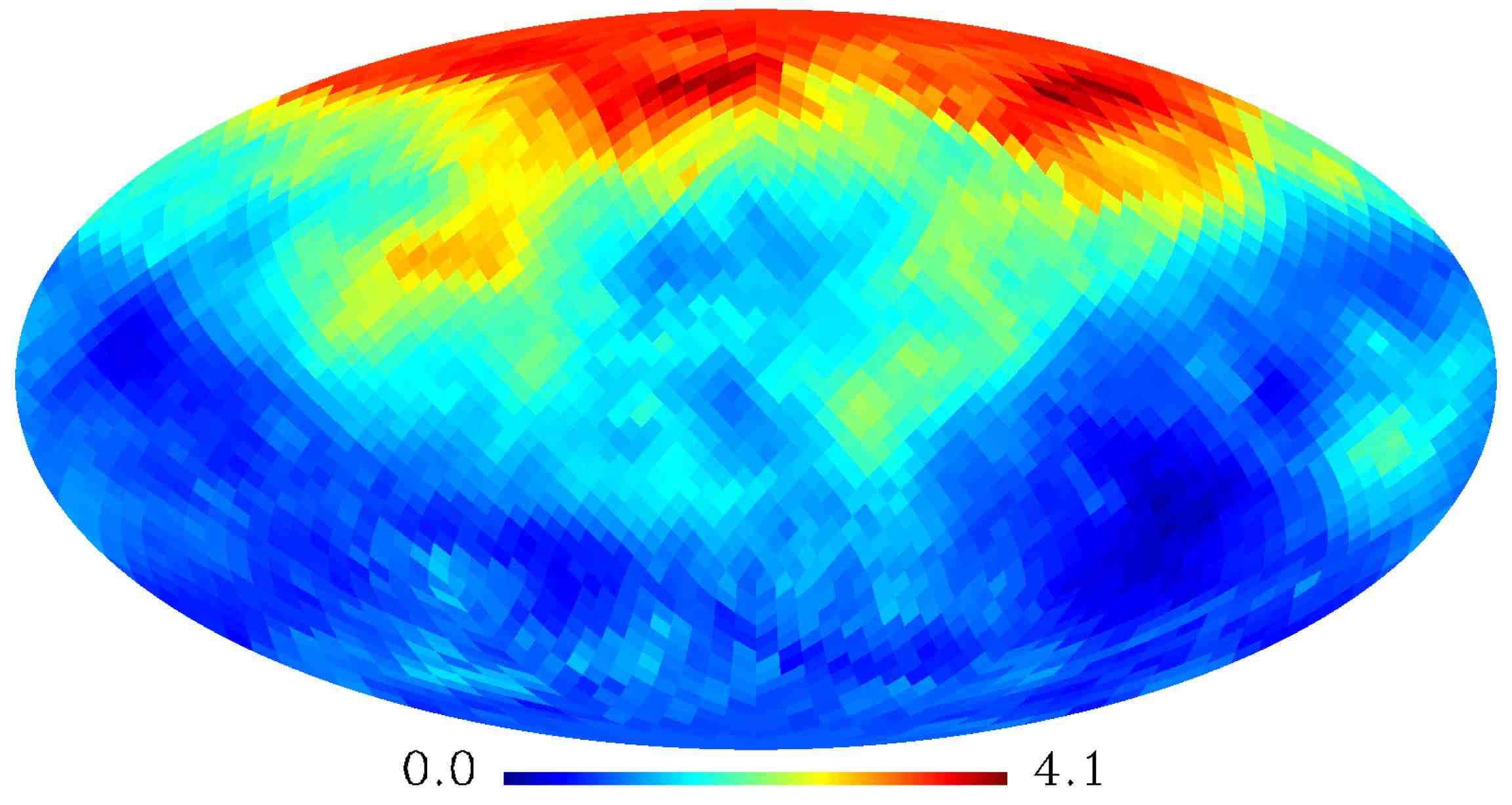}
\includegraphics[width=5.8cm, keepaspectratio=true]{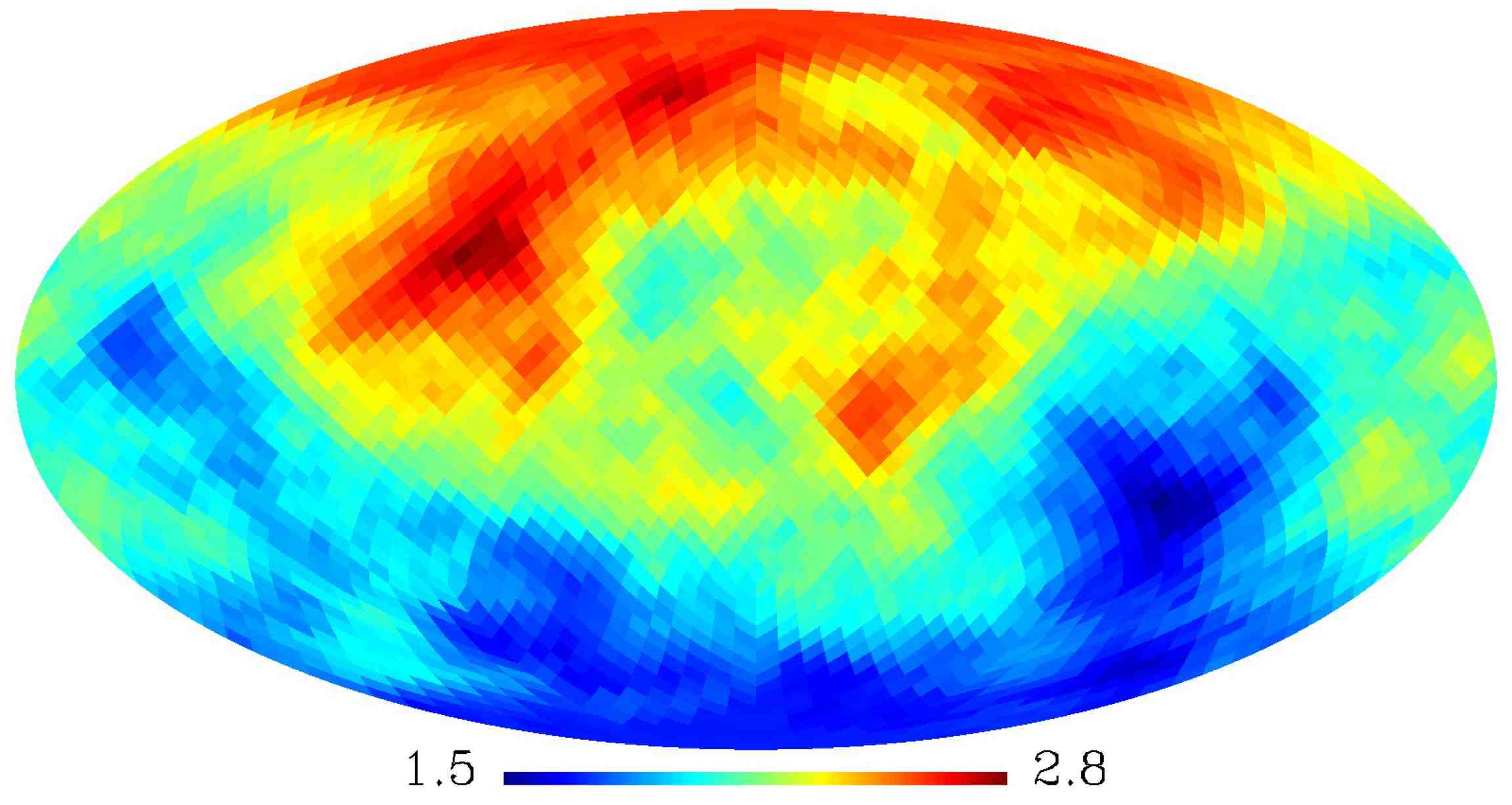}
\includegraphics[width=5.8cm, keepaspectratio=true]{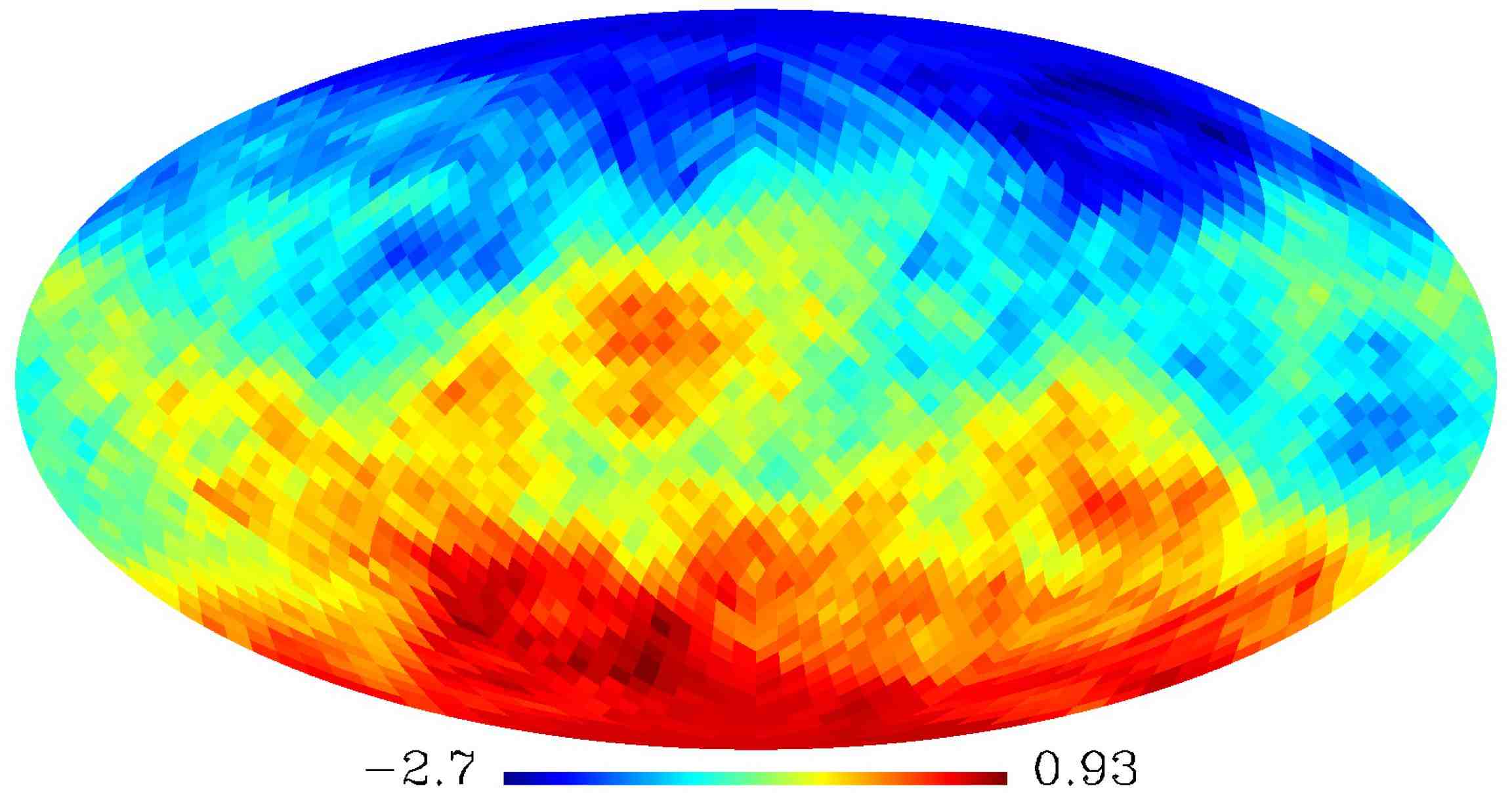}
\includegraphics[width=5.8cm, keepaspectratio=true]{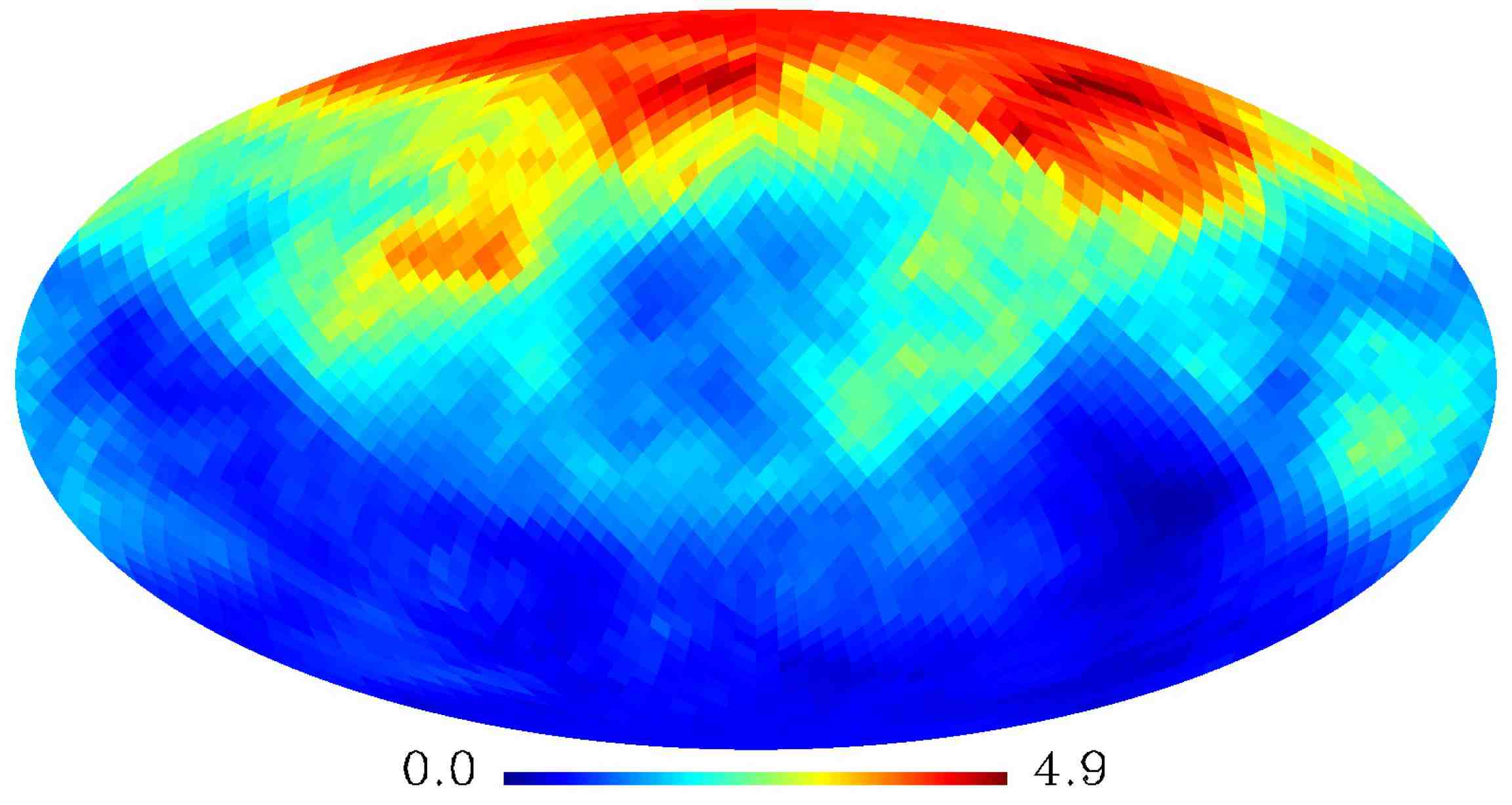}
\includegraphics[width=5.8cm, keepaspectratio=true]{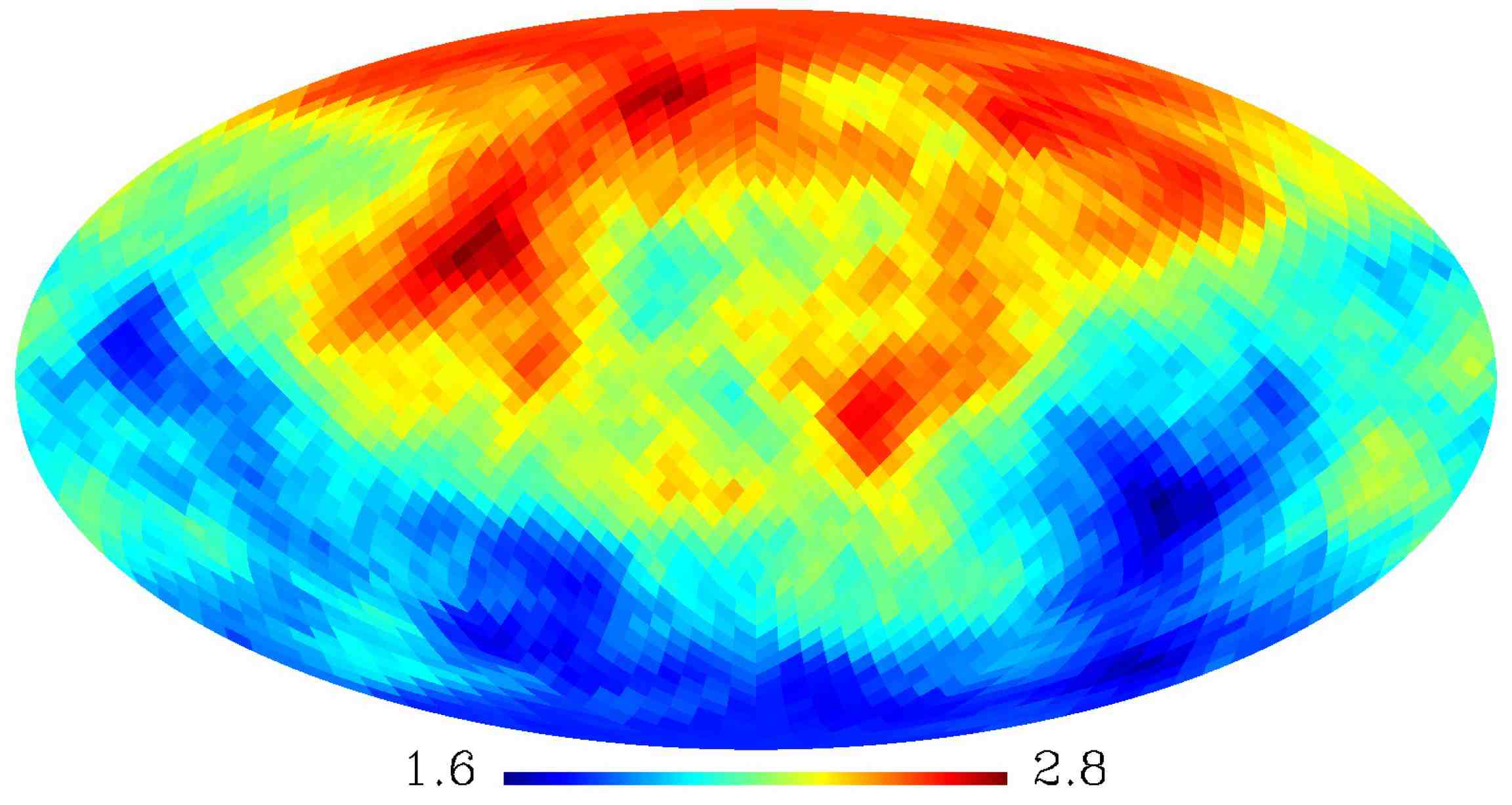}
\includegraphics[width=5.8cm, keepaspectratio=true]{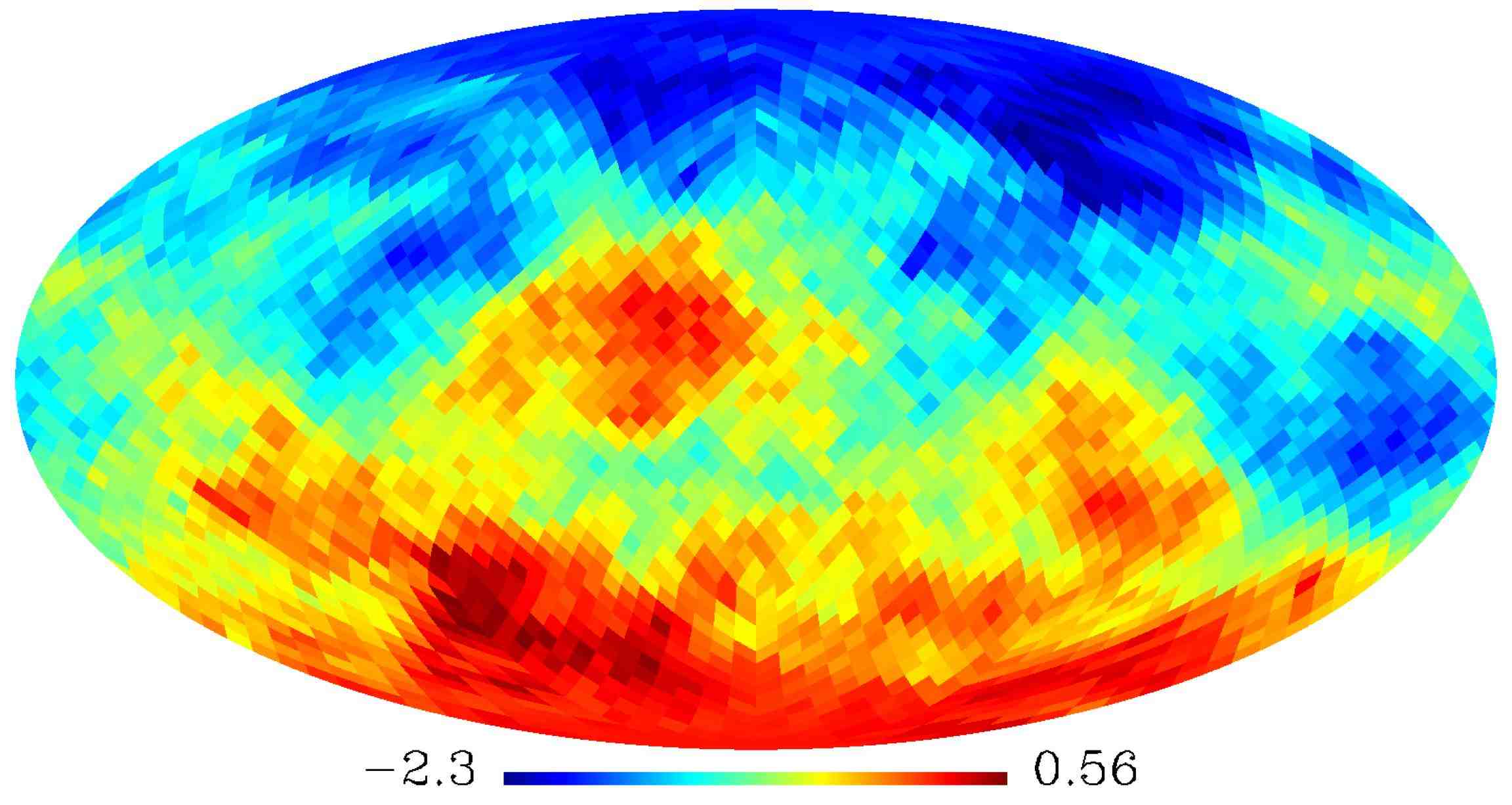}
\includegraphics[width=5.8cm, keepaspectratio=true]{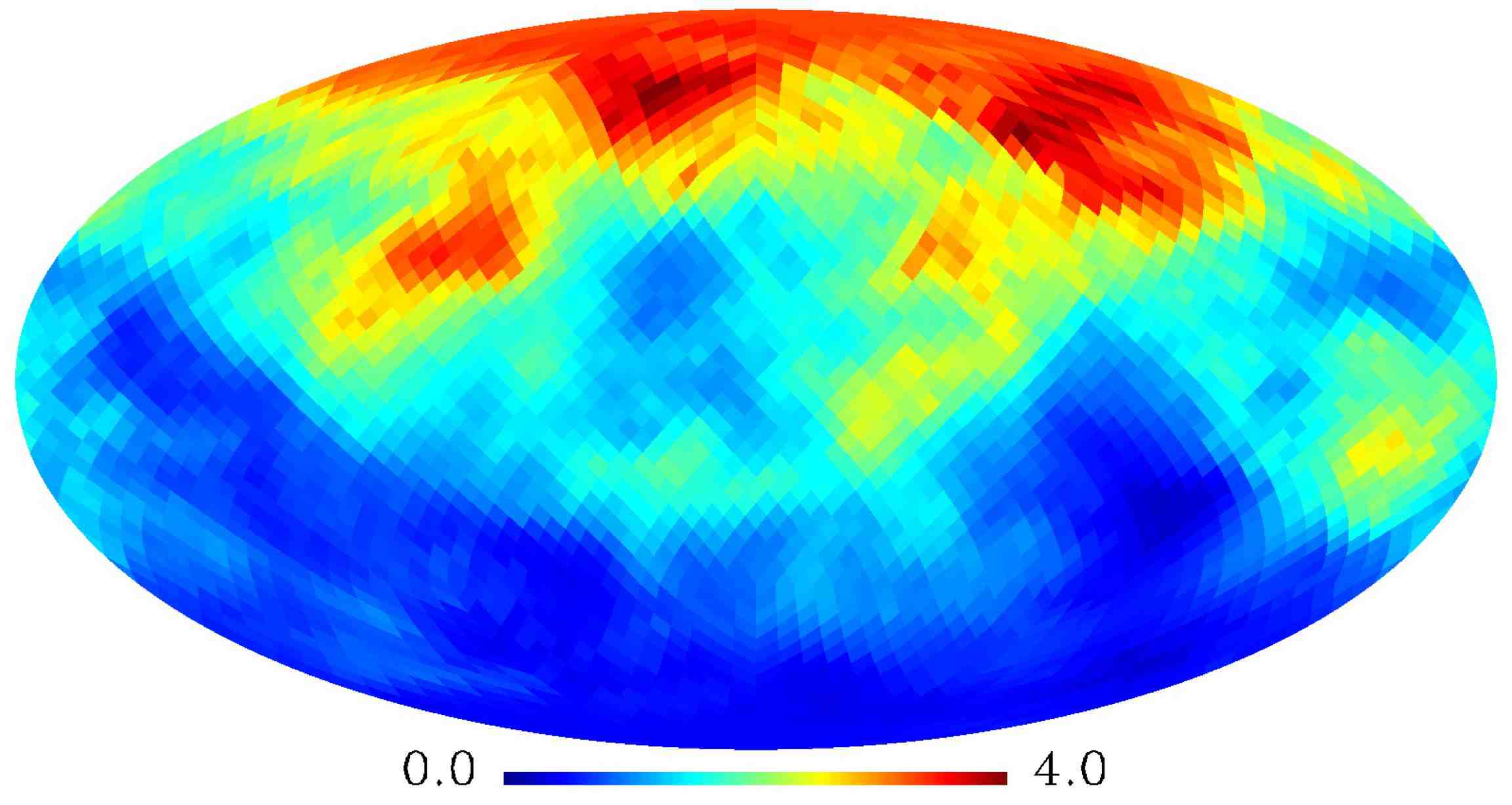}
\caption{The $\sigma$-normalised deviations $S(r)$ of the rotated hemispheres at the scale parameter $r=0.2$ for the mean (left column), the standard deviation (central column) and the diagonal $\chi^2$-statistics (right column) for the co-added VW-Band without (top row) and with (second row) the appliance of the mask filling method, as well as for the single Q-, V- and W-bands (third to fifth row), for which the mask filling method was always applied. Notice the different colour scaling for each plot.} \label{fig6:Rotationen}
\end{figure*}

Evidence for north-south asymmetry in the WMAP data was already detected using the angular power spectrum \citep{hansen04b, hansen08a} and higher order correlation functions \citep{eriksen04a}, spherical wavelets \citep{vielva04a}, local curvature analysis \citep{hansen04a}, two-dimensional genus measurements \citep{park04a} as well as all three Minkowski functionals \citep{eriksen04b}, correlated component analysis \citep{bonaldi07a}, spherical needlets \citep{pietrobon08a}, frequentist analysis of the bispectrum \citep{land05a}, two-point correlation functions \citep{bernui08a, bernui08b} and Bayesian analysis of the dipole modulated signal model \citep{hoftuft09a}. To take a closer look at asymmetries in the WMAP five-year data in our investigations, we perform an analysis of rotated hemispheres, as it was done for the three-year data in \citet{raeth07a}: For 3072 different angles, we rotate the original and simulated maps and then compute $S(r)$ for the above statistics (mean, standard deviation and $\chi^2$-combination) by only using the data in the resulting new upper hemisphere. Thus, the colour of each pixel in the corresponding figure \ref{fig6:Rotationen} expresses the positive or negative $\sigma$-normalised deviation $S(r)$ of the hemisphere around that pixel in the WMAP-data compared to the hemispheres around that pixel in the simulations. We apply this analysis for the co-added VW-band as well as for the single bands, whereas for the VW-band we use both the original and the mask filling method, but for the single bands the filling method only. In all charts of figure \ref{fig6:Rotationen} we can detect an obvious asymmetry in the data: The largest deviations between the data and the simulations are exclusively obtained for rotations pointing to northern directions relative to the galactic coordinate system. The maximum value for $S(r)$ of the $\chi^2$ analysis (right column of figure \ref{fig6:Rotationen}) using the mask-filling method on the co-added VW-band is obtained in the reference frame pointing to $(\theta,\phi) = (27^\circ,35^\circ)$, which is close to the galactic north pole. This proximity to the pole is consistent to the results of \citet{hansen04a} and \citet{raeth07a}, as well as to those findings of Hansen et al. (2004b) and \citet{eriksen04a} that considers large angular scales. For the standard deviation (central column of figure \ref{fig6:Rotationen}), the northern and southern hemispheres offer different algebraic signs. The negative $S(r)$ of the north implies a lower variability than the simulations in this region, while the south shows a converse behaviour. The fact that the plots using the new method show slightly lower values for $S(r)$ than the ones using the old method may be explained by the fraction of pure noise values within every rotated hemisphere, that diminish the degree of difference between the data and the simulations. \par

\begin{figure*}
\centering
\includegraphics[width=8cm, keepaspectratio=true]{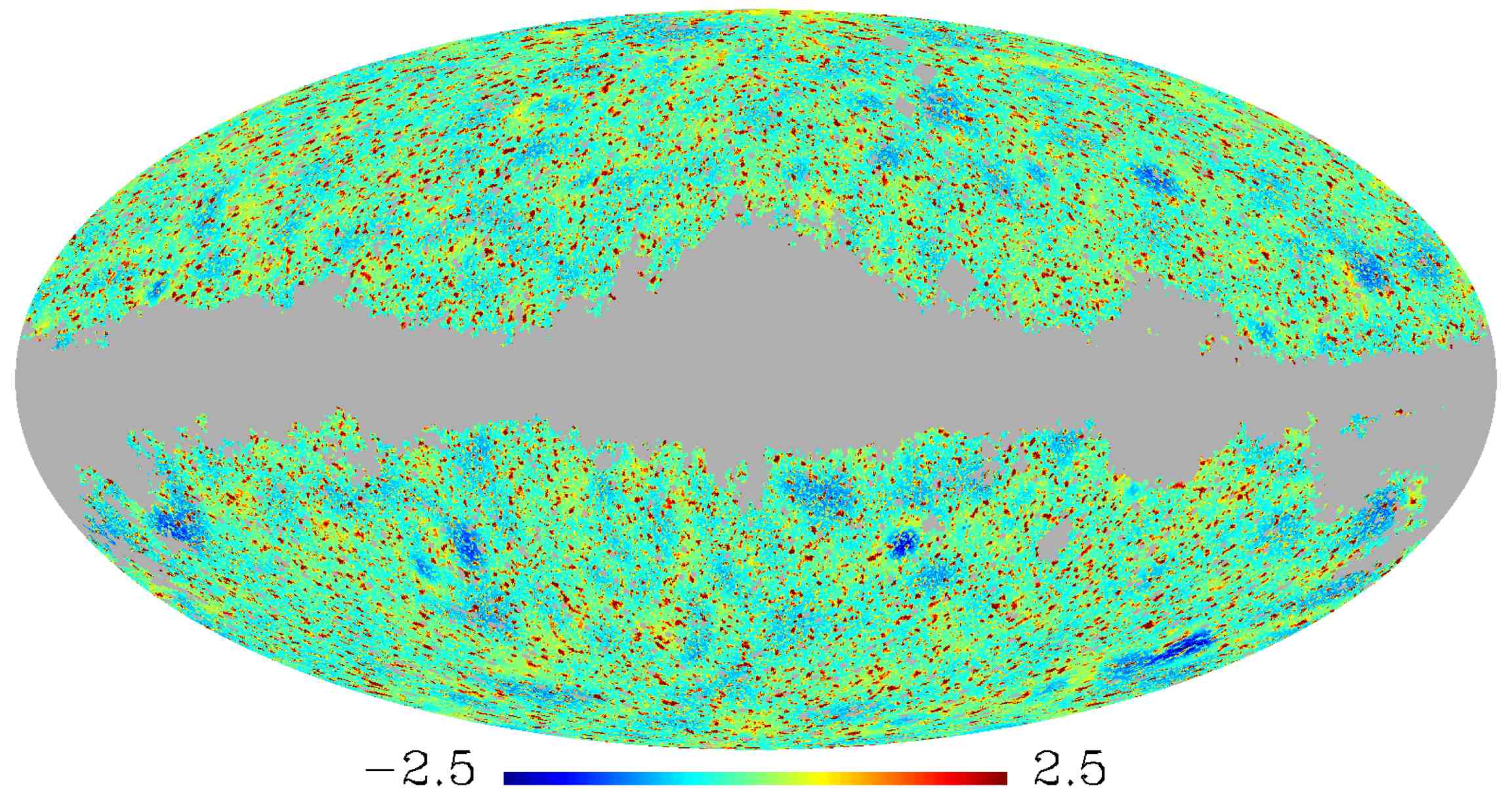}
\includegraphics[width=8cm, keepaspectratio=true]{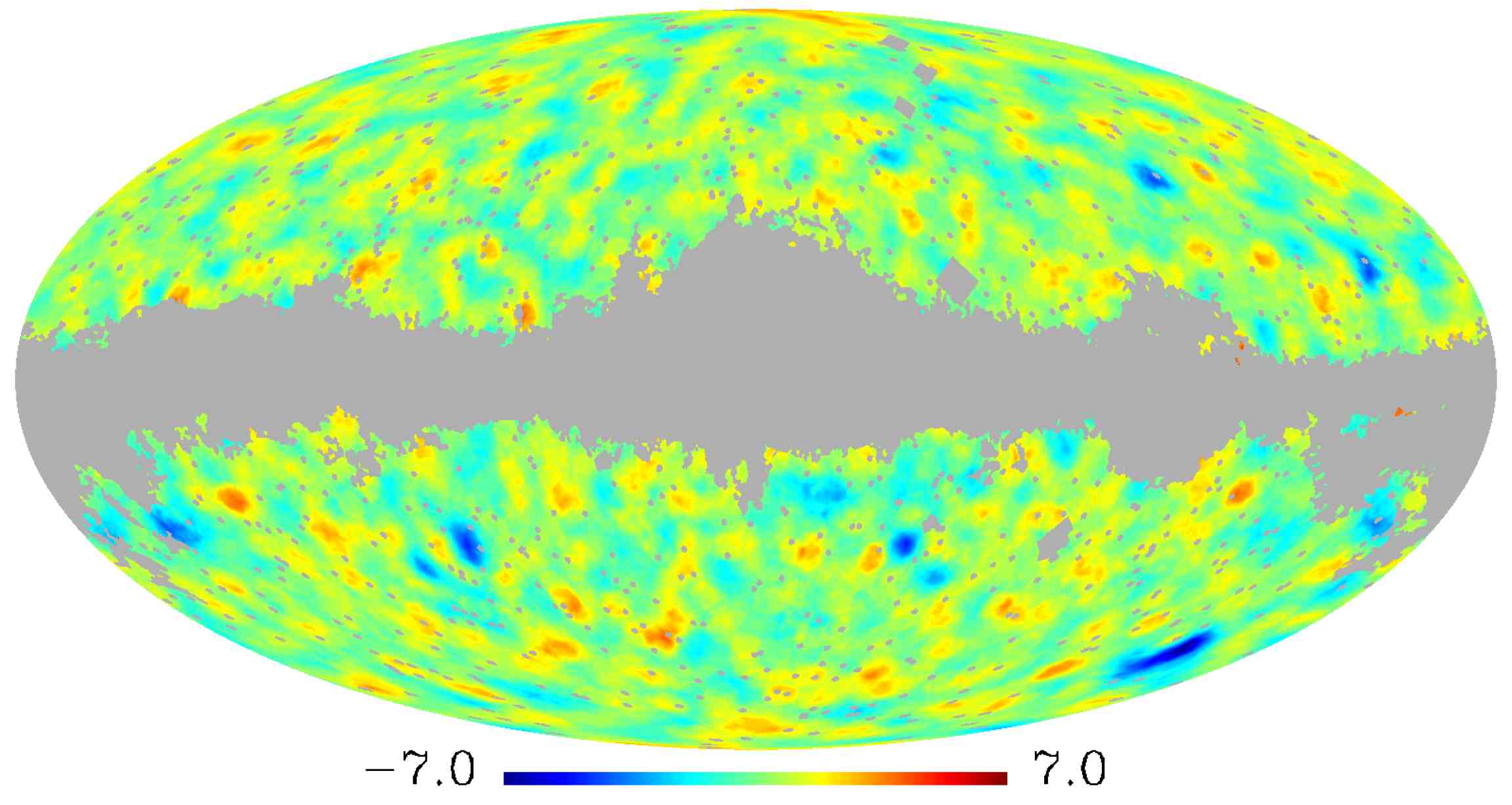}
\includegraphics[width=8cm, keepaspectratio=true]{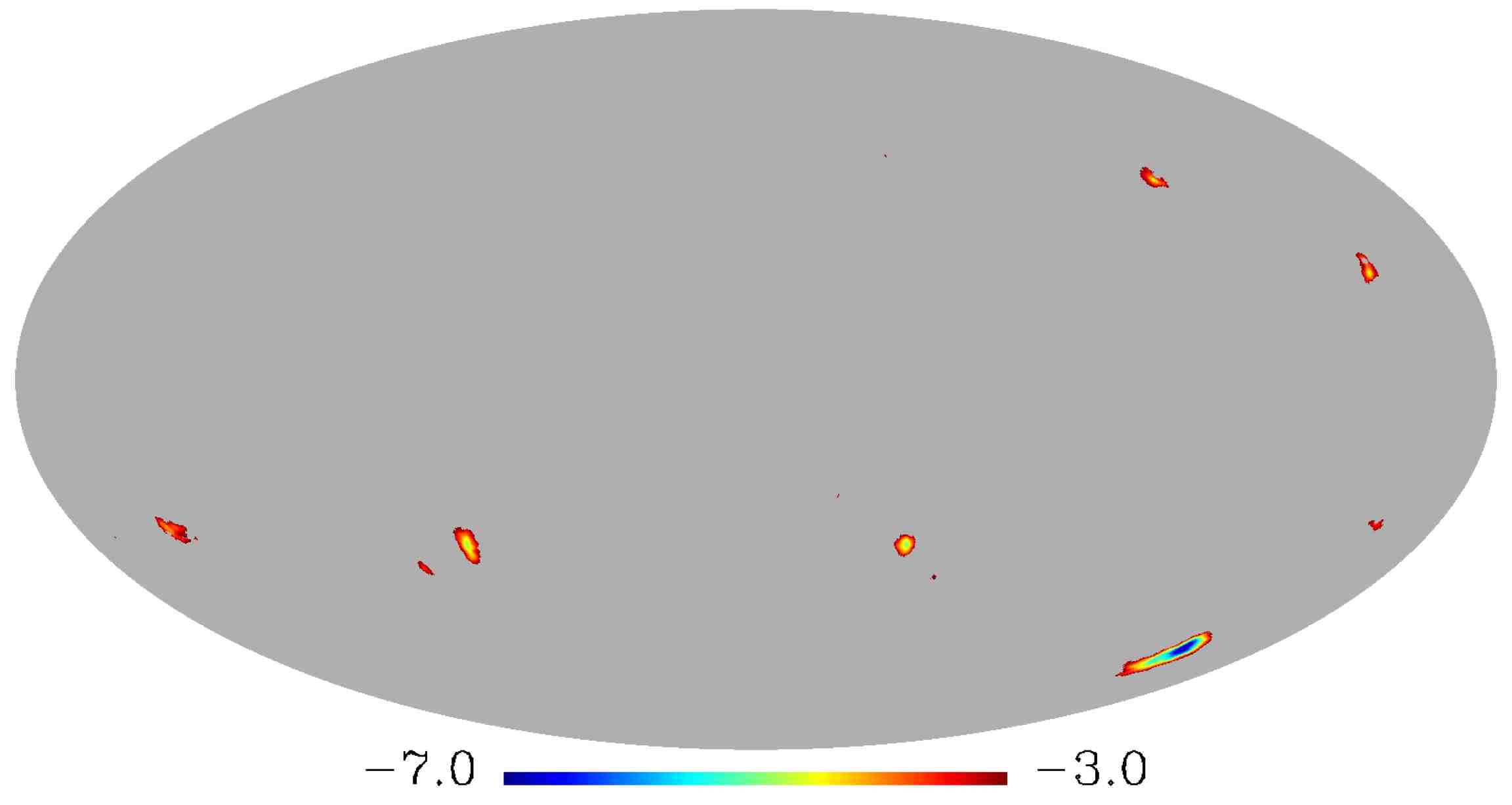}
\includegraphics[width=8cm, keepaspectratio=true]{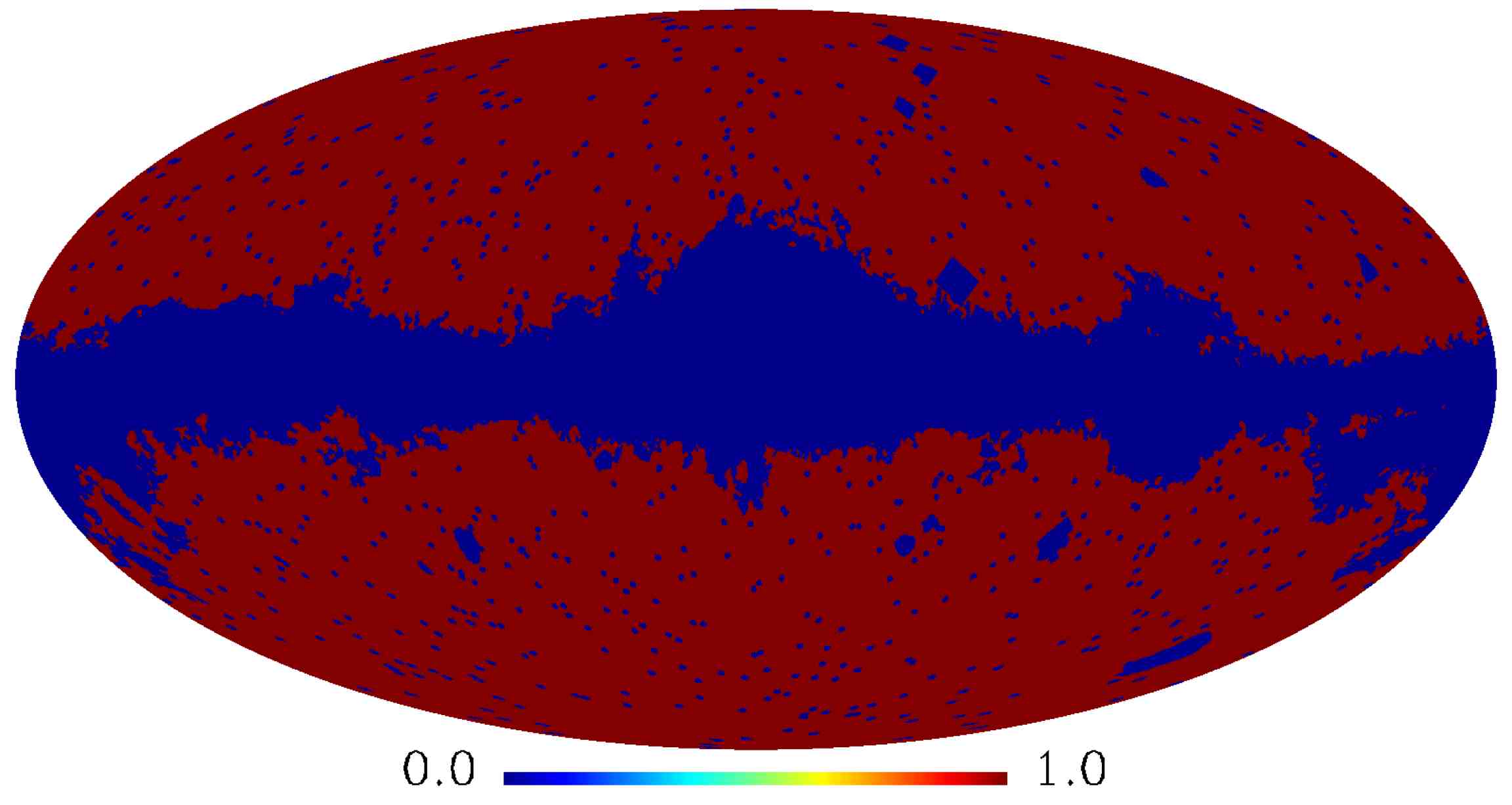}
\caption{The pixel-wise deviations $S_{i,r}$ of the primal (upper left) and of the smoothed scaling indices map (upper right), both based on the VW-band and the scale parameter $r=0.2$. The plot in the lower left only shows the values $\leq -3.0$ of the smoothed method. Except for the very small spots in the right part of this mapping, these regions are added to the KQ75-mask. The result is illustrated in the lower right plot.} \label{fig7:ColdSpots}
\end{figure*}

Another remarkable feature of figure \ref{fig6:Rotationen} is the high correlation between the different bands, that is visible to the naked eye but also confirmed mathematically: By calculating correlations $c$ among all combinations of those bands where the mask filling method was applied, we obtain for the mean $c \geq 0.99$ and for the standard deviation as well as the $\chi^2$-statistics $c \geq 0.95$. While the Q-band is heavily foreground-affected, first of all by synchroton radiation as well as radiation from electron-ion scattering ("free-free emission"), the W-band is mainly distorted by Dust emission. The V-band is affected by all these foregrounds, even though less than the other bands. As mentioned in chapter \ref{Kapitel2}, we use the foreground-reduced maps in our analysis, but one could still expect some small interferences. Despite the different influences on the different bands, we obtain the same signatures of non-Gaussianity in all single bands as well as in the co-added VW-band. Therefore we conclude that the measured asymmetry is very unlikely the result of a foreground influence but has to be concluded of thermal origin.


\subsection{Local features}
\label{localfeatures}

An interesting anomaly in the CMB data is that there are small regions which show very high or very low values in some local structure analysis. Vielva et al. (2004) detected the first of these regions, the well-known \textit{cold spot} at $(\theta,\phi) = (147^\circ,209^\circ)$ a few years ago by using a wavelet analysis. This Spot was re-detected several times using amongst others wavelet analysis \citep{mukherjee04a, cayon05a, cruz05a, cruz07b}, scaling indices \citep{raeth07a} or the Kolmogorov stochasticity parameter \citep{gurzadyan08a}. Furthermore, there have been some investigations which, in addition to the re-detection of the first spot, detected secondary spots via directional \citep{mcewen05a, mcewen06a, mcewen08a} or steerable wavelets \citep{vielva07a}, needlets \citep{pietrobon08a} and again the Kolmogorov stochasticity parameter \citep{gurzadyan09a}. These spots could be the result of some yet not fully understood physical process. For the $\textit{cold spot}$ lots of theories already exist which try to explain its origin by second-order gravitational effects \citep{tomita05a, tomita08a}, a finite universe model \citep{adler06a}, large dust-filled voids \citep{inoue06a, inoue07a, rudnick07a, granett08a}, cosmic textures \citep{cruz07a}, non-Gaussian modulation \citep{naselski07a}, topological defects \citep{battye08a}, textures in a brane world model \citep{cembranos08a} or an asymptotically flat Lemaître-Tolman-Bondi model \citep{garcia08a, masina08a}. \par
For our investigations concerning spots in the WMAP data we only use the mask-filling method of chapter \ref{Copingboundary} due to the reasons already explained above. We extend the analysis of scaling indices by applying two different approaches to detect anomalies: The first one is to calculate the $\sigma$-normalised deviation of \textit{every pixel} on the $\alpha$-response of the CMB map. For a given scale parameter $r$, this is achieved by comparing the scaling index $\alpha(\vec{p}_i,r)$ of each vector $\vec{p}_i, i=1,...,N_{pix}$, of the original data with the mean of the corresponding values $\alpha_{\ell}(\vec{p}_{i},r), \ell = 1,...,N_{sim}$, of the simulations depending on their standard deviation, where $N_{sim}$ denotes the number of the simulations. Formally, this reads as:
\begin{equation} \label{PixelwSignifikFormel}
S_{i,r} = \frac{\alpha(\vec{p}_i,r) - \mu_{i,r}}{\sigma_{i,r}},
\end{equation}
with
\begin{align*}
\mu_{i,r} &= \frac{1}{N_{sim}} \sum_{\ell=1}^{N_{sim}} \alpha_{\ell}(\vec{p}_{i},r) \\
\sigma^2_{i,r} &= \frac{1}{N_{sim}-1} \sum_{\ell=1}^{N_{sim}} \left( \alpha_{\ell}(\vec{p}_{i},r) - \mu_{i,r} \right)^2
\end{align*}
The results are illustrated in the upper left part of figure \ref{fig7:ColdSpots}. \par

\begin{figure}
\centering
\includegraphics[width=8cm, keepaspectratio=true]{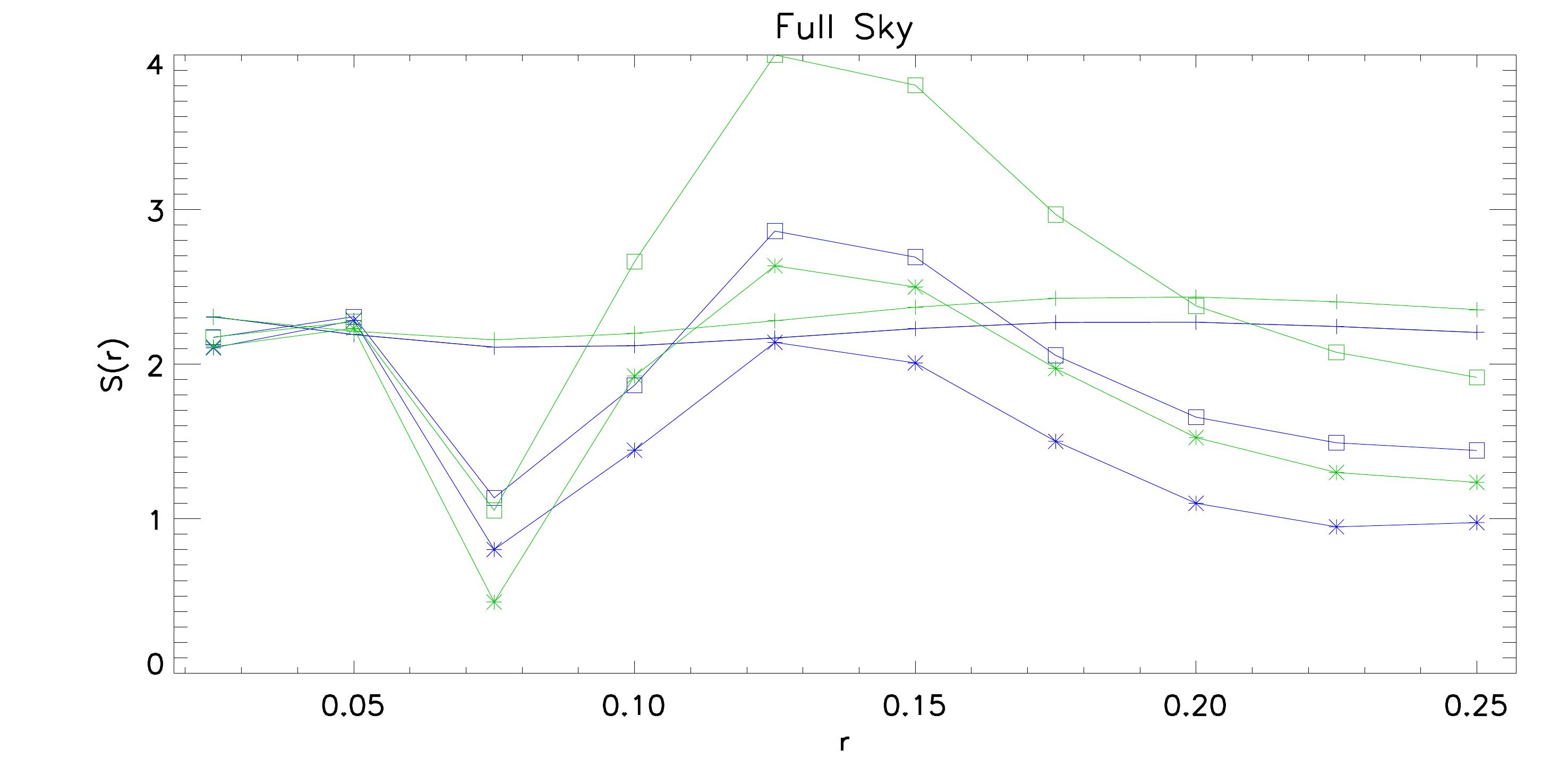}
\includegraphics[width=8cm, keepaspectratio=true]{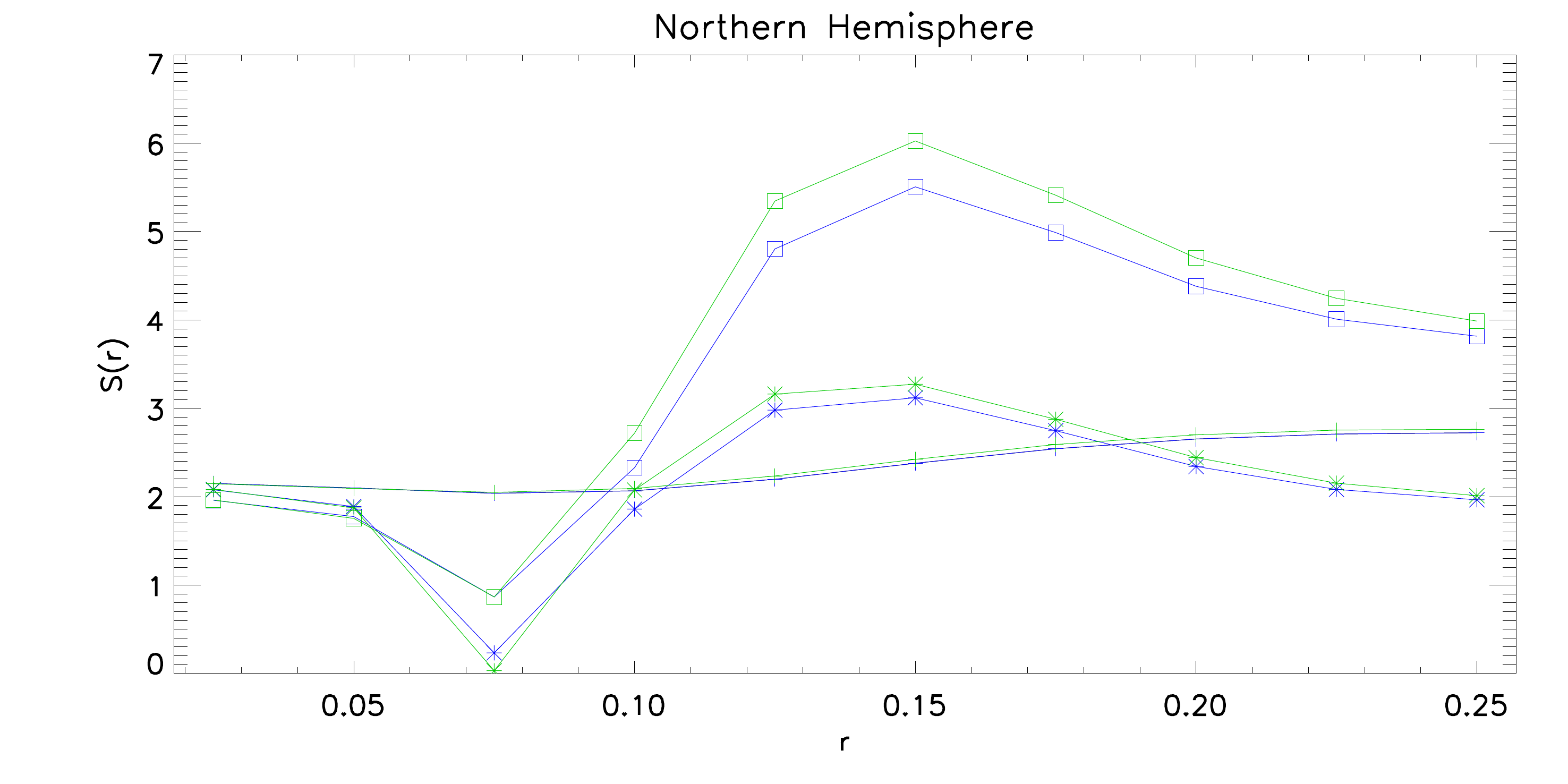}
\includegraphics[width=8cm, keepaspectratio=true]{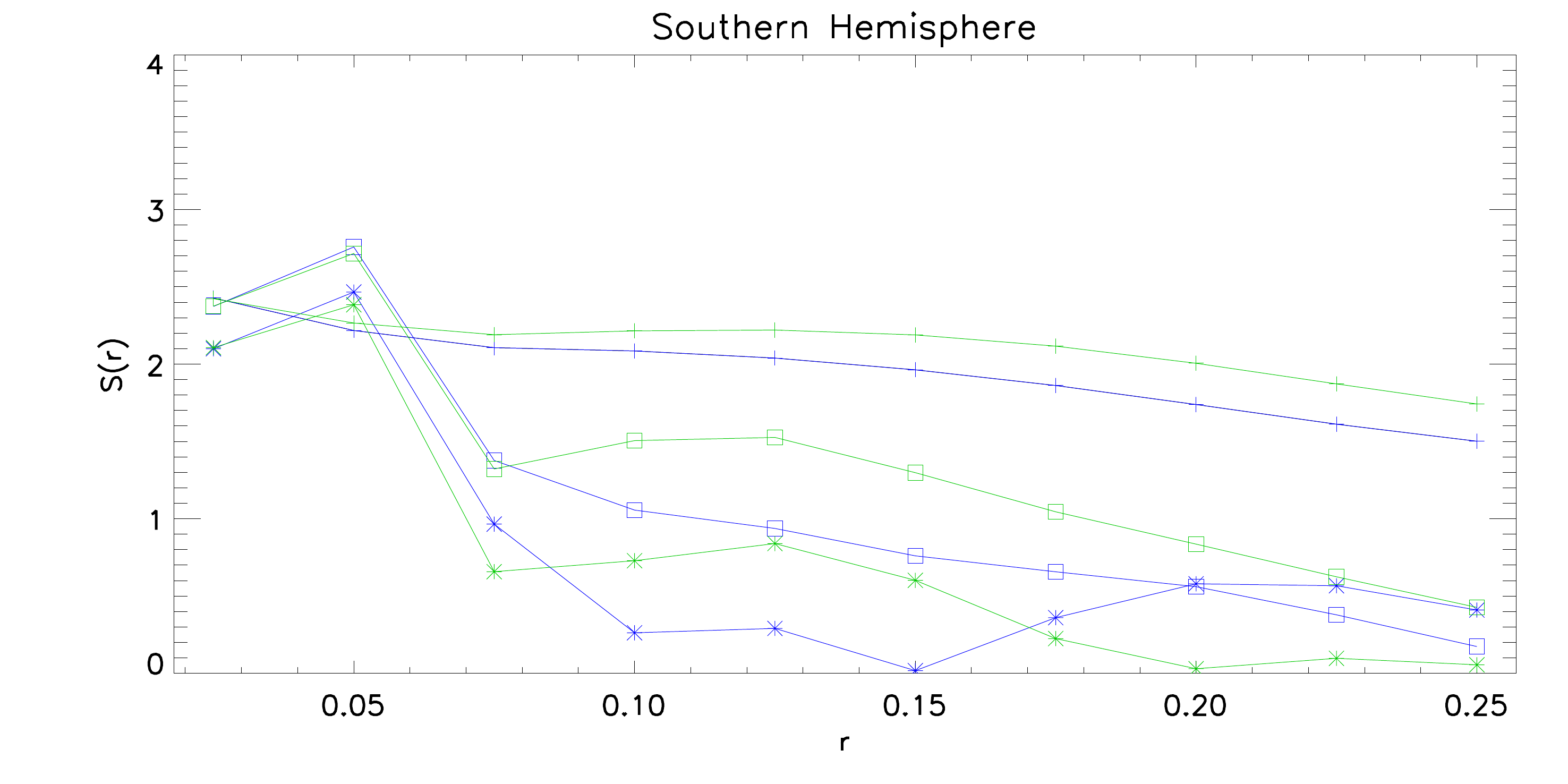}
\caption{The $\sigma$-normalised deviations of the mask-filling method for the original KQ75-mask (blue) and for the modified mask of the previous figure (green) in absolute values, plotted as a function of the scale parameter, whereby as above "$+$" denotes the mean, "$*$" the standard deviation and the boxes the $\chi^2$-combination. The full sky as well as again the single hemispheres are considered. The blue lines exactly correspond to the blue lines of figure \ref{fig4:SigVW}.} \label{fig8:SigVWnachMaske}
\end{figure}

\begin{figure}
\centering
\includegraphics[width=8cm, keepaspectratio=true]{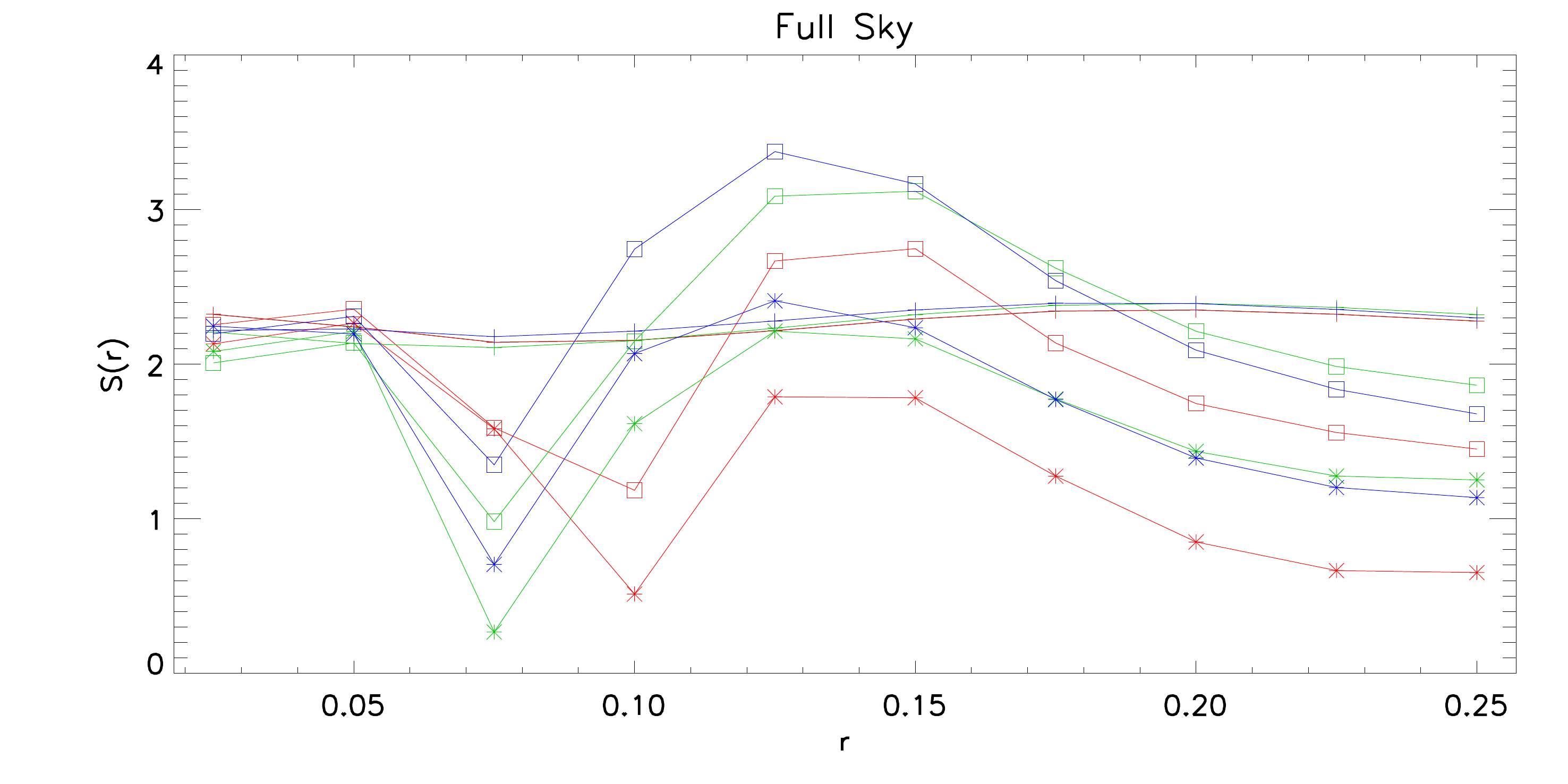}
\includegraphics[width=8cm, keepaspectratio=true]{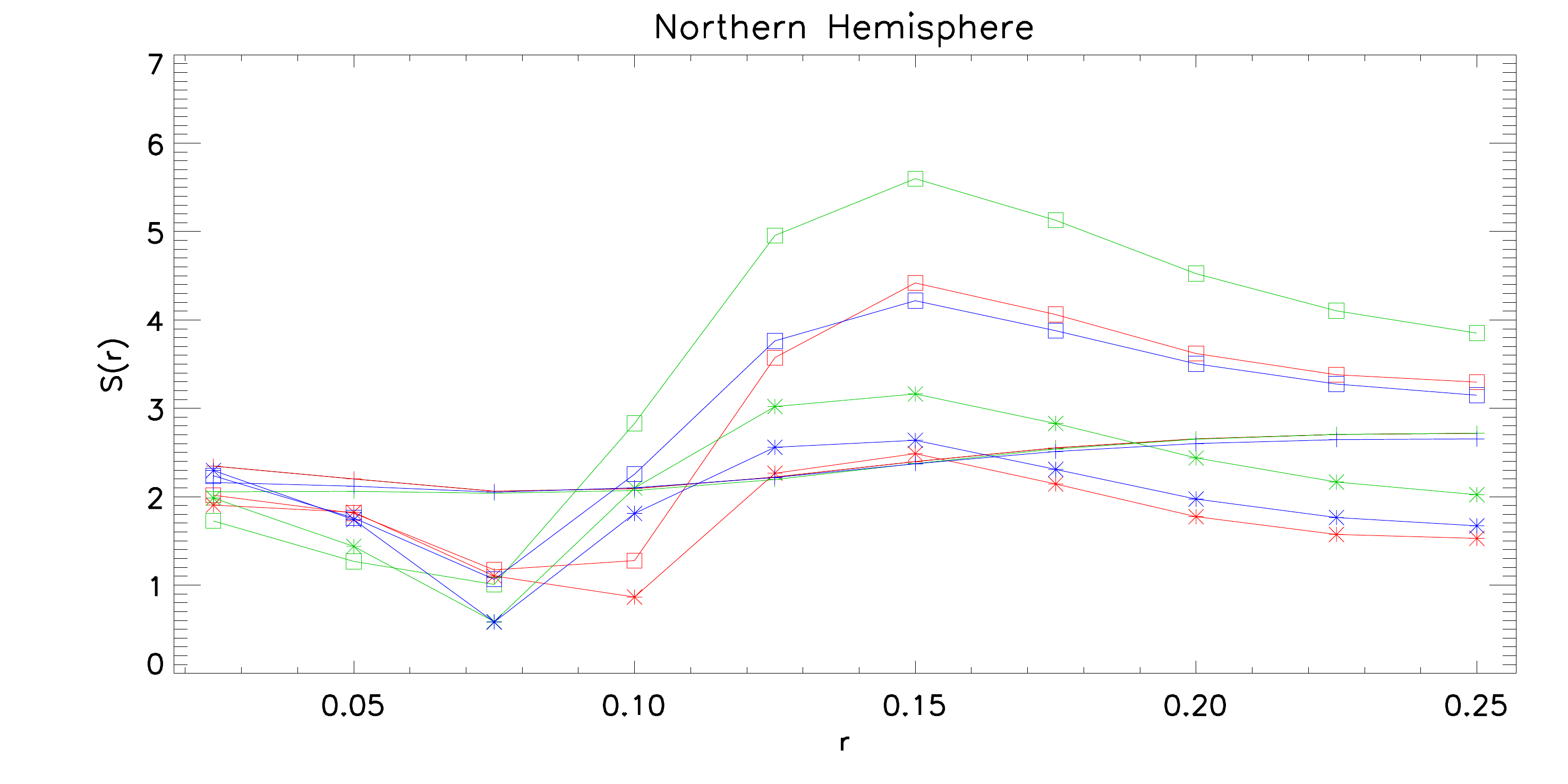}
\includegraphics[width=8cm, keepaspectratio=true]{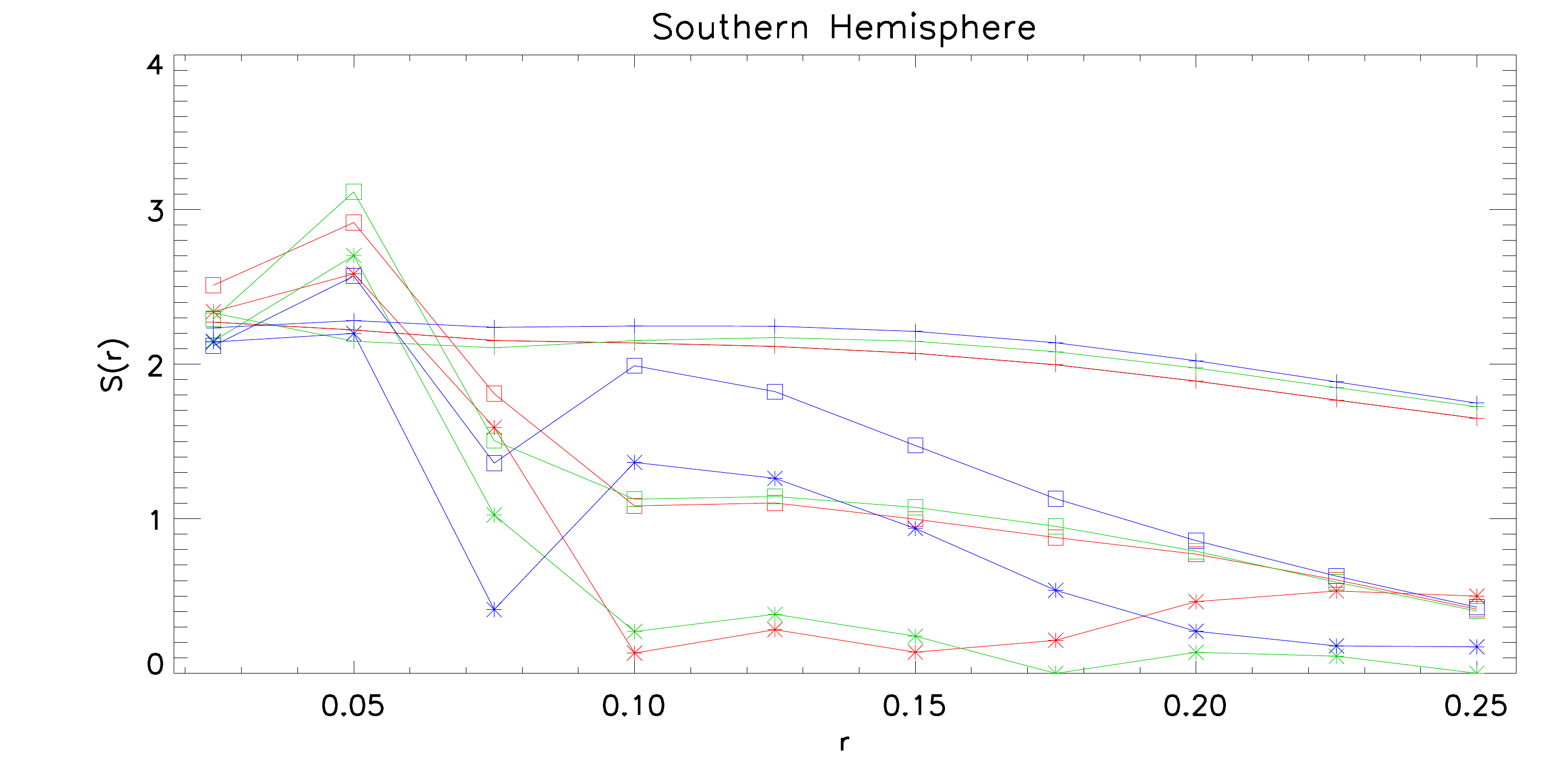}
\caption{Same as figure \ref{fig8:SigVWnachMaske} but exclusive for the modified mask of figure \ref{fig7:ColdSpots} applied to the Q- (red), V- (green) and W-band (blue). This plot is associated with figure \ref{fig5:SigQVW}.} \label{fig9:SigQVWnachMaske}
\end{figure}

The second approach smoothes the $\alpha$-maps of the original and simulated data by computing for every pixel the mean value of its surroundings given by some specified maximum distance, which equals $3^\circ$ in our analysis. We apply the pixel-wise deviations $S_{i,r}$ again on the resulting maps. The outcome of this procedure is shown in the upper right part of figure \ref{fig7:ColdSpots}. In the lower left plot of the same figure only the deviations $S_{i,r} \leq -3.0$ are illustrated to gain yet another clearer view on the interesting areas. \par
The first approach clearly shows the \textit{cold spot} and indicates some secondary spots in the southern as well as in the northern hemisphere. These get confirmed in the plot of the smoothing method, where we obtain a deviation of up to $-7\sigma$ for several clearly visible areas: In the southern hemisphere we detect a cold spot at $(\theta,\phi) = (124^\circ,320^\circ)$ and another one at $(\theta,\phi) = (124^\circ,78^\circ)$. Both were already detected with the above mentioned directional and steerable wavelet as well as with a needlet analysis. The former one is a \textit{hot} spot in these investigations. In our analysis, the latter spot actually appears as two spots close to each other, which is in agreement with \citet{pietrobon08a}. We discover another southern cold spot at $(\theta,\phi) = (120^\circ,155^\circ)$ which is very close to the mask. This spot represents a good example for the use of the mask filling method since it is situated at the edge of the non-masked region: The influence of the mask is diminishing the results of the calculation of the scaling indices in the area of this spot. This becomes obvious if one recalls the lower left plot of figure \ref{fig1:dreimaldrei}, in which the coordinates of the spot would be completely located in a "blue" region with low $\alpha$-values. Since the results of the scaling indices of local features show a similar, namely lower-valued, behaviour, an overlapping like that could prevent the detection of such spots close to the mask. By using the mask filling method, the detection of this cold spot on the edge of the mask is equivalent to a detection in an unmasked region, and therefore reliable. The spot at $(\theta,\phi) = (136^\circ,173^\circ)$, described by \citet{mcewen05a} and \citet{pietrobon08a}, is not traced in our analysis. In the northern hemisphere, our investigation shows two other cold spots at $(\theta,\phi) = (49^\circ,245^\circ)$ and $(\theta,\phi) = (68^\circ,204^\circ)$, which do not correspond with the so-called \textit{northern cold spot} of \citet{gurzadyan09a}, but with the results of \citet{mcewen05a}, where again one of them is a hot spot. Also \citet{pietrobon08a} locates one of these two spots. All these results were achieved with an analysis of the VW-band, but we find similar results in a single band analysis. \par
It is possible to define a new coordinate frame, including a new direction of the "north pole", such that all of these spots are contained in the "southern" hemisphere. This new north pole would then be located at $(\theta,\phi) = (51^\circ,21^\circ)$. \par
If the considered spots really depend on some yet not completely understood, maybe secondary, physical effect, they should not be implemented in a testing for intrinsic non-Gaussianity. For this reason, we modifiy the KQ75-mask by additionally excluding all above mentioned spots. A small peculiarity at the edge of the mask next to the \textit{cold spot} as well as three very small blurs in the right half of the lower left mollweide projection in figure \ref{fig7:ColdSpots} are not considered, since we regard their appearence as insufficent for being a distinctive feature. The modification of the KQ75-mask is illustrated in the lower right part of figure \ref{fig7:ColdSpots}. \par
We now apply this new mask to the $\alpha$-response of both the WMAP data as well as the simulations and repeat the analysis of chapter \ref{VierEins}. The results are illustrated in the figures \ref{fig8:SigVWnachMaske} and \ref{fig9:SigQVWnachMaske} as well as in table \ref{Table2}. A clear increase of $S(r)$ in comparison to the former analysis is evident. This heightening is in particular present in the southern hemisphere, where we detected more local features than in the north. The largest increase takes place in the co-added VW-band, where we now reach deviations of up to 4.0 for the $\chi^2$-combination in a full-sky analysis (former maximum: 2.9) and to the extend of 6.0 in an analysis of the northern hemisphere (former maximum: 5.5). But also the single bands in figure \ref{fig9:SigQVWnachMaske} as well as all scale-independent diagonal $\chi^2$-statistics in table \ref{Table2} show without exception a greater evidence for non-Gaussianity. \par

\begin{table}
\begin{tabular}{lcccc}
\hline \hline
 &  Full Sky  &  Northern Sky & Southern Sky \\ \hline
$\chi^2_{\langle \alpha \rangle}$: & $(S/\%)$ & $(S/\%)$ & $(S/\%)$ \\ \\
VW (mask-filled) & 2.4 / 97.6 & 2.8 / 98.4 & 2.0 / 97.2 \\
Q    (mask-filled) & 2.3 / 97.5 & 2.8 / 98.3 & 1.8 / 96.9 \\
V     (mask-filled) & 2.3 / 97.6 & 2.7 / 98.1 & 1.9 / 97.1 \\
W    (mask-filled) & 2.4 / 97.7 & 2.6 / 98.2 & 2.0 / 97.4 \\ \hline
$\chi^2_{\sigma_{\alpha}}$: & & & \\ \\
VW (mask-filled) & 2.6 / 96.7 & 4.8 / 99.8 & 0.2 / 97.2 \\
Q    (mask-filled) & 1.3 / 90.6 & 2.5 / 97.0 & 0.6 / 82.1 \\
V    (mask-filled) & 2.0 / 94.9 & 4.4 / 99.4 & 0.4 / 74.1 \\
W   (mask-filled) & 2.2 / 96.2 & 3.1 / 98.0 & 0.4 / 78.9 \\ \hline
$\chi^2_{\langle \alpha \rangle,\sigma_{\alpha}}$: & & & \\ \\
VW (mask-filled) & 2.7 / 98.0 & 4.0 / 99.1 & 1.6 / 96.2 \\
Q    (mask-filled) & 2.2 / 97.1 & 3.0 / 98.6 & 1.6 / 95.6 \\
V    (mask-filled) & 2.4 / 97.3 & 3.7 / 98.9 & 1.5 / 95.3 \\
W   (mask-filled) & 2.5 / 97.9 & 3.2 / 98.7 & 1.6 / 95.5 \\ \hline \hline
\end{tabular}
\caption{Same as table \ref{Table1}, but after excluding the cold spots via the modified KQ75-mask.} \label{Table2}
\end{table}

One could have expected to obtain higher values for $S(r)$ since the $\alpha$-response of the WMAP data in comparison to the one of the simulations featured a shift to higher values (see figure \ref{fig3:Overplots}): By now cutting out the local features, that exclusively consist of \textit{cold} spots in terms of pixel-wise deviations, one excludes spots that showed lower values than the average of the simulations (see equation \ref{PixelwSignifikFormel}). Therefore, the shift to higher values becomes even larger, hence leading to a higher $S(r)$. Still, the exclusion of the spots is helpful and necessary, since these local anomalies could origin in some independend physical process, as mentioned above.


\section{Summary}

We performed a scaling index analysis of the WMAP 5-year data following up the investigations of \citet{raeth07a}. For more realistic results around the mask, we additionally implemented a mask-filling method. By comparing the Q-,\linebreak[4] V-, W- and the co-added VW-band of the WMAP data with 1000 simulated maps per band, we (re)detected strong deviations from Gaussianity as well as asymmetries in the data, which can be summarized and interpreted as follows: \par
The scaling index values of the WMAP data are shifted to higher values and feature a higher variability than those of the simulations, especially in the northern hemisphere. This effect can be interpreted as less structure as well as more structural variations in the CMB signal compared to the corresponding Gaussian model. The results are confirmed by several statistics, that show deviations from Gaussianity of up to $5.9\sigma$ in the scale-dependent, and $5.5\sigma$ in the scale-independent case. These results are slightly lower applying the mask-filling method, and show high similarities within the different bands. In addition, we detected strong asymmetries by performing an analysis of rotated hemispheres: rotations pointing to northern directions show by far higher higher deviations from Gaussianity for the mean and the $\chi^2$-analysis than rotations pointing to the south. Observing the standard deviation, we obtained a negative outcome in the north and a positive in the south. This implies that the north possesses a more consistent pattern than the simulations, while the south shows the converse behaviour. This feature is in line with later investigations of local features, where we detected more local anomalies in the southern than in the northern hemisphere. \par
Furthermore, we performed an analysis of local features by studying pixel-wise deviations from Gaussianity with and without a previous smoothing of the $\alpha$-responses. For these investigations, we exclusively applied the mask filling method which can reduce the distorting effects on measurements like the scaling indices that appear when cutting out the masked regions. This mask-filling method eliminates the diluting effects on the border and therefore allows for an analysis of local features, which show a similar behaviour of lower outcome in the scaling index method. We detected the well-known \textit{cold spot} and three additional spots in the southern as well as two spots in the northern hemisphere. Except for one single spot in the south, all findings are in agreement with former results of different investigations. Since these spots could origin on some yet not completely understood physical effect, we excluded them from the data set and repeated the former analysis. Instead of obtaining lower deviations, the results show an increase of non-Gaussianity in all bands. Therefore, the discovered local anomalies are not the reason of the global detection of non-Gaussianity, but were actually dampening the deviations on average. In former isotropic wavelet-based analyses, an exclusion of detected spots lessened the significance level of indications of non-Gaussianity \citep{vielva04a}. Our new findings indicate in contrast, that the isotropic scaling index method can detect several different yet complementary aspects of the structural composition of the underlying data. The results of our investigation are in agreement with the steerable wavelet-based analysis in \citet{wiaux08a}, where the non-Gaussianites were conserved after excluding the detected local anomalies.


\section{Conclusions}

The redetection of indications for non-Gaussianity of the WMAP 3-year data analysis leads to the conclution that the observed results are not time-depending. In contrary, we can detect even higher deviations from the simulations which mimic the Gaussian properties of the best fit $\Lambda CDM$-model. Therefore, it is highly improbable  for the results to be caused by effects related to short-term measurements. \par
In addition, the coherence between the different analysed bands implies that  the foreground influence plays only a minor role but that the results are very unlikely to be truely of thermal origin. \par
Finally, the agreement of the detected spots with former investigations confirms the existence of these local anomalies.  \par
The two most important tasks for future studies are: First, to identify possible reasons for the indications of non-Gaussianity, which could be possible with the attainment of more and more precise data, e.g. with the upcoming PLANCK-mission. Second, to figure out possible sources of the observed local features and thereby solving the question, if these anomalies are due to systematics or foreground effects or indeed represent variations in the CMB signal itself.
 

\section*{Acknowledgments}

Many of the results in this paper have been derived using the HEALPix \citep{gorski05a} software and analysis package. We acknowledge use of the Legacy Archive for Microwave Background Data Analysis (LAMBDA). Support for LAMBDA is provided by the NASA Office of Space Science.


\label{lastpage}

\end{document}